\documentclass[12pt]{article}
\usepackage{lscape}
\usepackage{citesort}
\usepackage{amssymb}
\usepackage{appendix}
\usepackage{multirow}

\begin{document}

\def\qq{\langle \bar q q \rangle}
\def\uu{\langle \bar u u \rangle}
\def\dd{\langle \bar d d \rangle}
\def\sp{\langle \bar s s \rangle}
\def\GG{\langle g_s^2 G^2 \rangle}
\def\Tr{\mbox{Tr}}
\def\figt#1#2#3{
        \begin{figure}
        $\left. \right.$
        \vspace*{-2cm}
        \begin{center}
        \includegraphics[width=10cm]{#1}
        \end{center}
        \vspace*{-0.2cm}
        \caption{#3}
        \label{#2}
        \end{figure}
	}
	
\def\figb#1#2#3{
        \begin{figure}
        $\left. \right.$
        \vspace*{-1cm}
        \begin{center}
        \includegraphics[width=10cm]{#1}
        \end{center}
        \vspace*{-0.2cm}
        \caption{#3}
        \label{#2}
        \end{figure}
                }

\def\ds{\displaystyle}
\def\beq{\begin{equation}}
\def\eeq{\end{equation}}
\def\bea{\begin{eqnarray}}
\def\eea{\end{eqnarray}}
\def\beeq{\begin{eqnarray}}
\def\eeeq{\end{eqnarray}}
\def\ve{\vert}
\def\vel{\left|}
\def\ver{\right|}
\def\nnb{\nonumber}
\def\ga{\left(}
\def\dr{\right)}
\def\aga{\left\{}
\def\adr{\right\}}
\def\lla{\left<}
\def\rra{\right>}
\def\rar{\rightarrow}
\def\lrar{\leftrightarrow}  
\def\nnb{\nonumber}
\def\la{\langle}
\def\ra{\rangle}
\def\ba{\begin{array}}
\def\ea{\end{array}}
\def\tr{\mbox{Tr}}
\def\ssp{{\Sigma^{*+}}}
\def\sso{{\Sigma^{*0}}}
\def\ssm{{\Sigma^{*-}}}
\def\xis0{{\Xi^{*0}}}
\def\xism{{\Xi^{*-}}}
\def\qs{\la \bar s s \ra}
\def\qu{\la \bar u u \ra}
\def\qd{\la \bar d d \ra}
\def\qq{\la \bar q q \ra}
\def\gGgG{\la g^2 G^2 \ra}
\def\GG{\langle g_s^2 G^2 \rangle}
\def\g5{\gamma_5 \not\!q}
\def\x{\gamma_5 \not\!x}
\def\g5{\gamma_5}
\def\sb{S_Q^{cf}}
\def\sd{S_d^{be}}
\def\su{S_u^{ad}}
\def\sbp{{S}_Q^{'cf}}
\def\sdp{{S}_d^{'be}}
\def\sup{{S}_u^{'ad}}
\def\ssp{{S}_s^{'??}}

\def\sig{\sigma_{\mu \nu} \gamma_5 p^\mu q^\nu}
\def\fo{f_0(\frac{s_0}{M^2})}
\def\ffi{f_1(\frac{s_0}{M^2})}
\def\fii{f_2(\frac{s_0}{M^2})}
\def\O{{\cal O}}
\def\sl{{\Sigma^0 \Lambda}}
\def\es{\!\!\! &=& \!\!\!}
\def\ap{\!\!\! &\approx& \!\!\!}
\def\ar{&+& \!\!\!}
\def\arrr{\!\!\!\! &+& \!\!\!}
\def\ek{&-& \!\!\!}
\def\vev{&\vert& \!\!\!}
\def\kek{\!\!\!\!&-& \!\!\!}
\def\cp{&\times& \!\!\!}
\def\se{\!\!\! &\simeq& \!\!\!}
\def\eqv{&\equiv& \!\!\!}
\def\kpm{&\pm& \!\!\!}
\def\kmp{&\mp& \!\!\!}
\def\mcdot{\!\cdot\!}
\def\erar{&\rightarrow&}


\def\simlt{\stackrel{<}{{}_\sim}}
\def\simgt{\stackrel{>}{{}_\sim}}


\title{
         {\Large
                 {\bf
Diagonal and transition magnetic moments of negative parity heavy baryons
in QCD sum rules
                 }
         }
      }

\author{\vspace{1cm}\\
{\small T. M. Aliev$^1$ \thanks {e-mail: taliev@metu.edu.tr}~\footnote{permanent address:Institute of
Physics,Baku,Azerbaijan}\,\,,
{\small K. Azizi$^2$    \thanks {e-mail: kazizi@dogus.edu.tr}}\,\,,
T. Barakat$^3$ \thanks {e-mail:
tbarakat@KSU.EDU.SA}\,\,,
M. Savc{\i}$^1$ \thanks
{e-mail: savci@metu.edu.tr}} \\
{\small $^1$ Physics Department, Middle East Technical University,
06531 Ankara, Turkey }\\
{\small $^2$ Department of Physics, Do\u gu\c s University,
Ac{\i}badem-Kad{\i}k\"oy, 34722 \.{I}stanbul, Turkey}\\
{\small $^3$ Physics and Astronomy Department, King Saud University, Saudi
Arabia}}

\date{}

\begin{titlepage}
\maketitle
\thispagestyle{empty}

\begin{abstract}

Diagonal and transition magnetic moments of the negative parity, spin-1/2 heavy
baryons are studied in framework of the light cone QCD sum rules.
By constructing the sum rules for different Lorentz structures, the unwanted
contributions coming from negative (positive) to positive (negative) parity
transitions are removed. It is obtained that the magnetic moments of all
baryons, except $\Lambda_b^0$, $\Sigma_c^+$ and $\Xi_c^{\prime +}$, are
quite large. 
It is also found that the transition magnetic moments between
neutral negative parity heavy $\Xi_Q^{\prime 0}$ and $\Xi_Q^0$ baryons are
very small. Magnetic moments of the $\Sigma_Q \to \Lambda_Q$ and $
\Xi_Q^{\prime \pm} \to \Xi_Q^\pm$ transitions are quite large and can be
measured in further experiments.

\end{abstract}

~~~PACS numbers: 11.55.Hx, 13.40.Em, 14.20.Lq, 14.20.Mr

\end{titlepage}

\section{Introduction}

Last decade was quite fruitful in the field of heavy baryon spectroscopy.
At present, all baryons containing heavy charm and bottom quarks, except
$\Omega_b^\ast$ baryon, as well as several heavy baryons with negative
parity, have been observed by a number of collaborations (for a review see
\cite{Rsbs01}). These progresses stimulated further theoretical studies, and
experimental researches on heavy baryons at the existing facilities,
especially at LHC.

Heavy baryons are well recognized to represent a rich ``laboratory" for
theoretical investigations. The properties of heavy baryons have been
studied in framework of the various methods such as, relativistic quark
model \cite{Rsbs02}, variational approach \cite{Rsbs03}, constituent
quark model \cite{Rsbs04}, lattice QCD model \cite{Rsbs05}, QCD sum
rules method \cite{Rsbs06}. Recent progress on this subject can be found in
\cite{Rsbs07}.

Study of the electromagnetic properties of baryons
constitutes very important source of information in understanding their
internal structure, and can provide valuable insight about the
nonperturbative aspects of QCD. 
One of the most crucial quantities in this respect
is the magnetic moments of baryons. Magnetic moments of the ground
state $J^P = {1\over 2}^+$ heavy baryons have widely been studied
in literature. Furthermore, magnetic moments of the heavy baryons have been
investigated in the naive quark model in \cite{Rsbs08,Rsbs09},
phenomenological quark model \cite{Rsbs10}, relativistic three-quark model
\cite{Rsbs11}, variational approach \cite{Rsbs12}, nonrelativistic quark
model with screening and effective quark mass \cite{Rsbs13,Rsbs14},
nonrelativistic hypercentral model\cite{Rsbs15}, chiral constituent
quark model \cite{Rsbs16}, chiral bag model \cite{Rsbs17}, chiral
perturbation theory \cite{Rsbs18}, traditional QCD sum rules \cite{Rsbs19},
and light cone QCD sum rules (LCSR)
\cite{Rsbs20,Rsbs21,Rsbs22}, respectively.

In the present work, we calculate the diagonal and transition
magnetic moments of the negative parity,
spin-1/2 heavy baryons in framework of the light cone QCD sum rules method.

The body of the paper is organized as follows. In section 2, we construct
the light cone QCD sum rules for the diagonal and transition magnetic
moments of the negative parity, heavy baryons.
Section 3 is devoted to the numerical analysis of these sum rules
for the magnetic moments; and contains discussion of the
obtained results and conclusion.

\section{Sum rules for the diagonal and transition magnetic moments of
the negative parity heavy baryons}

In this section the LCSR for the negative parity baryons are derived. In
order to obtain the relevant sum rules, we consider the following
correlator,
\bea
\label{esbs01}
\Pi (p,q) = i \int d^4x e^{ipx} \lla 0 \vel \mbox{\rm T} \{
\eta_{Q_2} (x) \bar{\eta}_{Q_1}(0) \}\ver 0 \rra_\gamma ~,
\eea
where $\gamma$ means the external electromagnetic field, $\eta_Q (x)$ is the
interpolating current of the corresponding heavy baryon with spin-1/2. 
In order to obtain the sum rules for magnetic moments of the heavy baryons
the correlation function function is calculated in two different ways:
a) In terms of the hadrons; b) performing the operator product expansion (OPE) over
the twists of operators, and using the photon distribution amplitudes (DAs) which
encode all nonperturbative effects. The sum rules for the magnetic moments
of the heavy baryons are obtained by equating the coefficients of the
appropriate Lorentz structures that survive in parts (a) and (b). Finally,
in order to enhance the contributions coming from the ground states, and
suppress the contributions of the continuum and higher states, Borel
transformation with respect to the momentum squared of the initial and final
baryon states, and continuum subtraction procedure are performed,
successively (for more about the LCSR, see for example \cite{Rsbs23}).

To be able to calculate the correlator function in terms of the hadrons, we
insert the complete set of baryon states that carry the same quantum numbers
as the interpolating current $\eta(x)$. It should be noted here that the
interpolating current can produce both positive and negative parity baryons
from the vacuum state. Keeping this remark in mind, and isolating the
contributions of the ground state baryons, we get
\bea
\label{esbs02}
\Pi(p,q) \es
{\la 0 \ve \eta_{Q_2} \ve B_2^{(+)}(p,s) \ra \over p^2-m_{B_2^{(+)}}^2} \la B_2^{(+)}(p,s)
\gamma(q)\ve
B_1^{(+)}(p+q,s) \ra {\la B_1^{(+)}(p+q,s) \ve \bar{\eta}_{Q_1}(0) \ra \over
(p+q)^2-m_{B_1^{(+)}}^2} \nnb \\
\ar {\la 0 \ve \eta_{Q_2} \ve B_2^{(-)}(p,s) \ra \over p^2-m_{B_2^{(-)}}^2} \la B_2^{(-)}(p,s)
\gamma(q) \ve
B_1^{(-)}(p+q,s) \ra {\la B_1^{(-)}(p+q,s) \ve \bar{\eta}_{Q_1}(0) \ra \over
(p+q)^2-m_{B_1^{(-)}}^2} \nnb \\
\ar {\la 0 \ve \eta_{Q_2} \ve B_2^{(+)}(p,s) \ra \over p^2-m_{B_2^{(+)}}^2} \la B_2^{(+)}(p,s)
\gamma(q)\ve
B_1^{(-)}(p+q,s) \ra {\la B_1^{(-)}(p+q,s) \ve \bar{\eta}_{Q_1}(0) \ra \over
(p+q)^2-m_{B_1^{(-)}}^2} \nnb \\
\ar {\la 0 \ve \eta_{Q_2} \ve B_2^{(-)}(p,s) \ra \over p^2-m_{B_2^{(-)}}^2} \la B_2^{(-)}(p,s)
\gamma(q)\ve
B_1^{(+)}(p+q,s) \ra {\la B_1^{(+)}(p+q,s) \ve \bar{\eta}_{Q_1}(0) \ra \over
(p+q)^2-m_{B_1^{(+)}}^2} + \cdots~,
\eea
where $B^{(\pm)}$ and $m_{B^{(\pm)}}$ correspond to positive (negative)
parity baryons and their masses, respectively; $q$ is the photon momentum;
and dots correspond to the higher states contributions.

The matrix elements in Eq. (\ref{esbs02}) are defined as,
\bea
\label{esbs03}
\la 0 \ve \eta_{Q_2} \ve B^{(+)}(p)\ra \es  \lambda_{B^{(+)}} u^{(+)} (p)~,\nnb \\
\la 0 \ve \eta_{Q_2} \ve B^{(-)}(p)\ra \es  \lambda_{B^{(-)}} \gamma_5 u^{(-)} (p)~,\nnb \\
\la B_2^{(+)}(p) \gamma(q) \ve B_1^{(+)}(p+q)\ra \es e \varepsilon^\mu
\bar{u}^{(+)}(p)
\left[\gamma_\mu f_1 - {i \sigma_{\mu\nu} q^\nu \over  m_{B_1^{(+)}}
+ m_{B_2^{(+)}} } f_2 \right] u^{(+)}(p+q) \nnb \\
\es e \varepsilon^\mu \bar{u}^{(+)}(p) \left[ (f_1+f_2) \gamma_\mu - {(2p+q)_\mu
\over m_{B_1^{(+)}} + m_{B_2^{(+)}} } f_2 \right] u^{(+)}(p+q)~, \nnb \\
\la B_2^{(-)}(p) \gamma(q) \ve B_1^{(+)}(p+q)\ra \es e \varepsilon^\mu
\bar{u}^{(-)}(p)
\left[\gamma_\mu f_1^T - {i \sigma_{\mu\nu} q^\nu \over 
m_{B_1^{(+)}} + m_{B_2^{(-)}} } f_2^T
\right] \gamma_5 u^{(+)}(p+q) \nnb \\
\es e \varepsilon^\mu \bar{u}^{(-)}(p)
\left[\left( f_1^T - {m_{B_1^{(+)}} - m_{B_2^{(-)}} \over 
m_{B_1^{(+)}} + m_{B_2^{(-)}} } f_2^T \right) \gamma_\mu \right. \nnb \\
\ek \left. {(2p+q)_\mu \over m_{B_1^{(+)}} + m_{B_2^{(-)}} } f_2^T \right] \gamma_5
u^{(+)}(p+q)~, \nnb \\ 
\la B_2^{(-)}(p) \gamma(q) \ve B_1^{(-)}(p+q)\ra \es e \varepsilon^\mu
\bar{u}^{(-)}(p)
\left[\gamma_\mu f_1^\ast - {i \sigma_{\mu\nu} q^\nu \over 
m_{B_1^{(-)}} + m_{B_2^{(-)}} } f_2^\ast
\right] u^{(-)}(p+q) \nnb \\
\es e \varepsilon^\mu \bar{u}^{(-)}(p) \left[\left(f_1^\ast + f_2^\ast
\right) \gamma_\mu - {(2p+q)_\mu \over m_{B_1^{(-)}} + m_{B_2^{(-)}} }
f_2^\ast \right] u^{(-)}(p+q)~,
\eea    
where $\varepsilon^\mu$ is the four-polarization vector of the photon, and
$\lambda_{B^{(+)}}$,  $\lambda_{B^{(-)}}$ are the residues of the
positive and negative parity baryons, respectively.

Performing summation over spins of the
heavy baryons, for the correlation function from the hadronic side we get,
\bea
\label{esbs04} 
\Pi (p,q) \es A ({\not\!{p}}_2 +  m_{B_2^{(+)}}) 
\not\!{\varepsilon} ({\not\!{p}}_1 +  m_{B_1^{(+)}})~\nnb \\
\ar B ({\not\!{p}}_2 -  m_{B_2^{(-)}})       
\not\!{\varepsilon} ({\not\!{p}}_1 -  m_{B_1^{(-)}})~\nnb \\
\ar C  ({\not\!{p}}_2 -  m_{B_2^{(-)}})       
\not\!{\varepsilon} \gamma_5({\not\!{p}}_1 +  m_{B_1^{(+)}})~\nnb\\
\ar D  ({\not\!{p}}_2 +  m_{B_2^{(+)}})       
\not\!{\varepsilon} \gamma_5({\not\!{p}}_1 -  m_{B_1^{(-)}})~,
\eea
where
\bea
\label{esbs05}
A\es { \lambda_{B_1^{(+)}}  \lambda_{B_2^{(+)}} (f_1+f_2) \over
( m_{B_1^{(+)}}^2-p_1^2)( m_{B_2^{(+)}}^2-p_2^2)} \nnb \\
B\es { \lambda_{B_1^{(-)}}  \lambda_{B_2^{(-)}} (f_1^\ast+f_2^\ast) \over
( m_{B_1^{(-)}}^2-p_1^2)( m_{B_2^{(-)}}^2-p_2^2)} \nnb \\
C\es { \lambda_{B_1^{(-)}} \lambda_{B_2^{(+)}} \over
( m_{B_1^{(-)}}^2-p_1^2)( m_{B_2^{(+)}}^2-p_2^2)}
\left[f_1^T + { m_{B_1^{(-)}} -  m_{B_2^{(+)}} \over m_{B_1^{(-)}} + m_{B_2^{(+)}}} f_2^T \right]\nnb \\
D\es { \lambda_{B_1^{(+)}} \lambda_{B_2^{(-)}} \over
( m_{B_1^{(+)}}^2-p_1^2)( m_{B_2^{(-)}}^2-p_2^2)}
\left[f_1^T - { m_{B_1^{(+)}} - m_{B_2^{(-)}} \over   m_{B_1^{(+)}} + m_{B_2^{(-)}} } f_2^T \right]~,
\eea
where $p_1 = p+q$ and $p_2=p$.

Among the terms in Eq. (\ref{esbs04})
\bea
\label{nolabel01}
f_1+f_2~,~~(f_1^\ast+f_2^\ast)~,~~f_1^T + { m_{B_1^{(-)}} - m_{B_2^{(+)}}
\over  m_{B_1^{(-)}} + m_{B_2^{(+)}} } f_2^T~,~~f_1^T - { m_{B_1^{(+)}} -
m_{B_2^{(-)}} \over m_{B_1^{(+)}} + m_{B_2^{(-)}} } f_2^T~,\nnb
\eea
that are proportional to $\gamma_\mu$, the first two correspond
to the magnetic moments of the positive to
positive, negative to negative transitions, respectively;
and the third and the fourth ones describe the transition magnetic moments between positive and
negative parity baryons at $q^2=0$.
Our aim in the present work
is to calculate the diagonal and the transition magnetic moment between the negative parity
baryons, and therefore we should find a way to remove the other three contributions.

In order to determine the diagonal and the transition magnetic moments between negative
parity baryons four equations are needed, for which we choose the following
four Lorentz structures, $(\varepsilon\!\cdot\! p) I$,
$(\varepsilon\!\cdot\! p) \rlap/{p}$, $\rlap/{p}\rlap/{\varepsilon}$ and
$\rlap/{\varepsilon}$. Solving finally these four coupled equations, we
obtain the unknown coefficient $B$ which describes the negative to negative
parity transition.

It follows from Eq. (\ref{esbs01}) that interpolating currents are needed
in order to calculate the correlation function in terms of quarks and
gluons. Here, it should be remembered that hadrons containing single heavy
quark belong to either sextet or anti-triplet representations of $SU(3)$.
Sextet (anti-triplet) representation is symmetric (antisymmetric) with
respect to the exchange of light quarks. The heavy baryons $\Sigma_Q$,
$\Xi_Q^\prime$, and $\Omega_Q$ belonging to the sextet; and $\Xi_Q$ and
$\Lambda_Q$ belonging to the triplet representations of the $SU(3)$ group.
Using this fact the general form of interpolating
currents belonging to sextet and anti-triplet representations can be 
written in the following form (see \cite{Rsbs24}),
\bea
\label{esbs06}
\eta^{(s)} \es -{1\over \sqrt{2}} \varepsilon^{abc} \Big\{
(q_1^{aT} C Q^b) \gamma_5 q_2^c +
t (q_1^{aT} C \gamma_5 Q^b) q_2^c +
(q_2^{aT} C Q^b) \gamma_5 q_1^c + (q_2^{aT} C
\gamma_5 Q^b) q_1^c\Big\}~, \nnb \\
\eta^{(a)} \es -{1\over \sqrt{6}} \varepsilon^{abc} \Big\{
2 (q_1^{aT} C q_2^b) \gamma_5 Q^c +
2 t (q_1^{aT} C \gamma_5 q_2^b) Q^c +
(q_1^{aT} C Q^b) \gamma_5 q_2^c \nnb \\
\ar t (q_1^{aT} C \gamma_5 Q^b) q_2^c -
(q_2^{aT} C Q^b) \gamma_5 q_1^c -
t (q_2^{aT} C \gamma_5 Q^b) q_1^c\Big\}~.
\eea
where $t$ is an arbitrary parameter ($t=-1$ corresponds to the
so-called Ioffe current); $a,b,c$ are the color indices; and $C$ is the
charge conjugation operator.
The light quark contents of the sextet and antitriplet representations are
given in Table 1.


\begin{table}[h]

\renewcommand{\arraystretch}{1.3}
\addtolength{\arraycolsep}{-0.5pt}
\small
$$
\begin{array}{|c|c|c|c|c|c|c|c|c|c|c|}
\hline \hline   
 & \Sigma_{c(b)}^{(++)+} & \Sigma_{c(b)}^{+(0)} & \Sigma_{c(b)}^{0(-)} &
             \Xi_{c(b)}^{\prime 0(-)}  & \Xi_{c(b)}^{\prime +(0)}  &
 \Omega_{c(b)}^{0(-)} & \Lambda_{c(b)}^{+(0)} & 
\Xi_{c(b)}^{0(-)}  & \Xi_{c(b)}^{+(0)} \\   
\hline \hline
q_1   & u & u & d & d & u & s & u & d & u \\
q_2   & u & d & d & s & s & s & d & s & s \\
\hline \hline 
\end{array}
$$
\caption{Quark contents of the heavy baryons belonging to the
sextet and antitriplet representations.}
\renewcommand{\arraystretch}{1}
\addtolength{\arraycolsep}{-1.0pt}
\end{table}      


Using the interpolating currents given in Eq. (\ref{esbs06}),
one can easily calculate the theoretical part of the
correlator function. As an example, we present the form of the correlation
function for $\Sigma_Q^+$ in terms of the corresponding propagators,
\bea
\label{esbs07}
\Pi^{\Sigma_Q^+} \es
-2 \varepsilon^{abc} \varepsilon^{a^\prime b^\prime c^\prime} \int d^4x \la
0 \ve \sum_{\ell=1}^2 \sum_{k=1}^2 \Big\{
A_2^\ell S_u^{cc^\prime} (x) A_2^k \mbox{Tr} S_Q^{bb^\prime} (x) C A_1^k
S_u^{aa^\prime} (x) C A_1^\ell \nnb \\
\ar A_2^\ell \Big[ S_u^{cc^\prime} (x) (C A_1^k)^T S_Q^{bb^\prime T} (x)
(C A_1^\ell)^T S_u^{aa^\prime} (x) A_2^k \nnb \\
\ar S_u^{aa^\prime} (x) (C A_1^k)^T S_Q^{bb^\prime T} (x)
(C A_1^\ell)^T S_u^{cc^\prime} (x) A_2^k \nnb \\
\ar S_u^{aa^\prime} (x) A_2^k \mbox{Tr} S_Q^{bb^\prime T} (x)
C A_1^k S_u^{cc^\prime T} (x) C A_1^\ell \Big] \Big\} \ve 0 \ra_\gamma~,
\eea
where $A_1^1=1$, $A_1^2=t \gamma_5$, $A_2^1=\gamma_5$, $A_2^2=t$,
and $S_q(x)$ and $S_Q(x)$ are the full propagators of the light and heavy
quarks.

The expressions of the correlator functions for the $\Sigma_Q^-$,
$\Sigma_Q^0$, $\Xi_Q^{\prime 0}$, $\Xi_Q^{\prime -}$ and
$\Omega_Q$ can be found by performing the following replacements,
\bea
\label{esbs08}   
\Pi^{\Sigma_Q^-} \es \Pi^{\Sigma_Q^+} (u\to d)~, \nnb \\
\Pi^{\Sigma_Q^0} \es {1\over 2} \left( \Pi^{\Sigma_Q^+} +
\Pi^{\Sigma_Q^-}\right)~, \nnb \\
\Pi^{\Xi_Q^0} \es \Pi^{\Sigma_Q^+} (u\to s)~, \nnb \\
\Pi^{\Xi_Q^-} \es \Pi^{\Sigma_Q^+} (u\to s, s\to d)~.
\eea

As has already been noted, in the present work we also calculate the magnetic
moments of $\Sigma_Q \to \Lambda_Q$ and $\Xi_Q^\prime \to \Xi_Q$
transitions. It follows from Eq. (\ref{esbs07}) in order to calculate the
correlation function from the QCD side, the expressions of the light and
heavy quark propagators in the presence of the external field are needed. 
The light cone expression of the light quark propagator in external field is
calculated in \cite{Rsbs25} in which it is found that the contributions of
the nonlocal operators such as $\bar{q} G q$, $\bar{q} G^2 q$,
$\bar{q} q \bar{q} q$, are small. Neglecting these operators is justified in
conformal spin expansion \cite{Rsbs26}. Note that in further analysis we
retain only those terms that are linear in quark mass. The expression of the
light quark operator in the presence of external field is given as,
\bea
\label{esbs09}
S_q(x) \es {i \rlap/x\over 2\pi^2 x^4} - {m_q\over 4 \pi^2 x^2} -
{\lla \bar q q \rra\over 12} \left(1 - i {m_q\over 4} \rlap/x \right) -
{x^2\over 192} m_0^2 \lla \bar q q \rra  \left( 1 -
i {m_q\over 6}\rlap/x \right) \nnb \\
\ek i g_s \int_0^1 du \Bigg[{\rlap/x\over 16 \pi^2 x^2} G_{\mu \nu} (ux)
\sigma_{\mu \nu} - {i\over 4 \pi^2 x^2} u x^\mu G_{\mu \nu} (ux) \gamma^\nu \nnb \\
\ek i {m_q\over 32 \pi^2} G_{\mu \nu} \sigma^{\mu
 \nu} \left( \ln {-x^2 \Lambda^2\over 4}  +
 2 \gamma_E \right) \Bigg] + \cdots~,
\eea
where $\Lambda$ is the cut-off energy separating perturbative and
nonperturbative domains, and $\gamma_E$ is the Euler constant.

In calculating the correlation function from the QCD side we also need the
expression for the heavy quarks, whose explicit form in the coordinate space
can be expressed as,
\bea
\label{esbs10}
S_Q(x) \es {m_Q^2 \over 4 \pi^2} \Bigg\{ {K_1(m_Q\sqrt{-x^2}) \over \sqrt{-x^2}} +
i {\rlap/{x} \over -x^2} K_2(m_Q\sqrt{-x^2}) \Bigg\} \nnb \\ 
\ek {g_s \over 16 \pi^2} \int_0^1 du
G_{\mu\nu}(ux) \left[ \left(\sigma^{\mu\nu} \rlap/x + \rlap/x
\sigma^{\mu\nu}\right) {K_1 (m_Q\sqrt{-x^2})\over \sqrt{-x^2}} +
2 \sigma^{\mu\nu} K_0(m_Q\sqrt{-x^2})\right] +\cdots~,
\eea
where $K_i(m_Q\sqrt{-x^2})$ are the modified Bessel functions.
Taking into account the expressions of the light and heavy propagators, the
correlation function given in Eq. (\ref{esbs08}) can be calculated from the
QCD side. This correlation function contains
three different type of contributions: a)
Perturbative contributions, i.e., photon interacts with the quark
propagators perturbatively. Technically this contribution can be calculated
by replacing the one of the free quark operators (the first two terms in
Eqs. (\ref{esbs09}) and (\ref{esbs10})) by,
\bea
\label{esbs11}
S^{free} (x) \to \int d^4y S^{free} (x-y) \not\!\!{A} (y)  S^{free} (x-y)~,
\eea
and the remaining two propagators are the free ones. b) In the case when
photon interacts with the heavy quark perturbatively, the free part must be
removed at least in one of the light quark propagators, i.e., one (or
both)light quark operators are replaced by the quark condensate.
c) Nonperturbative
contributions, i.e., photon interacts with the light quark fields at large
distance. This contribution can be calculated by replacing one of the light
quark operators by,
\bea
\label{esbs12}
S_{\alpha\beta}^{ab} \to -{1\over 4} \left( q^a \Gamma_i q^b \right)
\left(\Gamma_i \right)_{\alpha\beta},
\eea
and the remaining quarks constitute the full quark propagators. Here,
$\Gamma_j$ are the full set of Dirac matrices
$\gamma_j=\left\{I,\gamma_5,\gamma_\mu,i
\gamma_\mu\gamma_5,\sigma_{\mu\nu}/\sqrt{2} \right\}$. When Eq.
(\ref{esbs12}) is used in calculation of the nonperturbative contributions,
we see that matrix elements of the form $\la \gamma(q) \ve \bar{q} \Gamma_i
q \ve 0 \ra$ are needed. These matrix elements are defined in terms of the
photon distribution amplitudes \cite{Rsbs27},and are presented in Appendix A.

As has already been noted, in determination of the magnetic moment
responsible for the negative to negative parity transition, four equations
are needed, and for this purpose we choose the coefficients of the
structures $(\varepsilon\!\cdot\! p) I$,
$(\varepsilon\!\cdot\! p) \rlap/{p}$, $\rlap/{p}\rlap/{\varepsilon}$ and  
$\rlap/{\varepsilon}$. The sum rules for the negative to negative parity
transition magnetic moments can be obtained by choosing the coefficients of
the aforementioned Lorentz structures $\Pi_i$, and equate them to the
corresponding coefficients in hadronic part. Solving then the linear
equations for the coefficients describing the negative to negative
transition magnetic moments, and performing Borel transformation over the
variables $-p^2$ and $-(p+q)^2$ in order to suppress higher states and
continuum contribution, we finally obtain the magnetic moment for the negative
to negative parity baryon transitions as is given below,
\bea                                                                        
\label{esbs13}
\mu \es {e^{m_{B_1^{(-)}}^2/2M^2} e^{m_{B_2^{(-)}}^2/2M^2}  \over
2 \lambda_{B_1^{(-)}}  \lambda_{B_2^{(-)}}
\left( m_{B_1^{(+)}} +  m_{B_1^{(-)}}\right)
\left( m_{B_2^{(+)}} +  m_{B_2^{(-)}}\right)}
\Big\{ \left( m_{B_1^{(+)}} + m_{B_2^{(-)}}\right)
\left(\Pi_1^{B} - m_{B_2^{(+)}} \Pi_2^{B}\right) \nnb \\
\ar 2 m_{B_2^{(+)}} \Pi_3^{B} - 2 \Pi_4^{B}\Big\}~.
\eea
In this expression we take $M_1^2=M_2^2=2 M^2$, since the masses of the
diagonal transitions are same, and masses of the baryons that responsible
for the 
$\Sigma_Q \to \Lambda_Q$ and  $\Xi_Q^\prime \to \Xi_Q$ transitions
baryons are very close
to each other. The expressions of $\Pi_i^B$ for the $\Sigma_b^0$ baryon
and $\Xi_b^{\prime 0} \to \Xi_b^0$ transition are presented in Appendix B.

It follows from Eq. (\ref{esbs13}) that in determination of the diagonal and
transition magnetic
moments, the residues  of the heavy negative parity, spin-1/2 baryons 
are needed.
These residues can be determined from the analysis of the
two-point correlation function
\bea
\label{nolabel06} 
\Pi(q^2)= i \int d^4x e^{iqx} \la 0 \ve \mbox{\rm T} \left\{ \eta_Q(x)
\bar{\eta}_Q(0) \right\} \ve 0 \ra~, \nnb
\eea
where $\eta_Q$ is the interpolating current for the corresponding heavy
baryon given by Eq. (\ref{esbs06}). This interpolating current interacts
with both positive and negative parity heavy baryons. Saturating this
correlation function with the ground states of positive and negative parity
baryons we have,
\bea
\label{nolabel07}
\Pi (q^2) = {\ve \lambda_{B^{(-)}} \ve^2 (\not\!p - m_{B^{(-)}}) \over
m_{B^{(-)}}^2-p^2} +
{\ve \lambda_{B^{(+)}} \ve^2 (\not\!p + m_{B^{(+)}}) \over m_{B^{(+)}}^2-p^2}~\nnb .
\eea
Eliminating the contributions coming from the positive parity baryons, the
following sum rules for the residue and mass of the negative parity baryons
are obtained,
\bea
\label{nolabel08}
\ve \lambda_{B^{(-)}} \ve^2 \es {1\over \pi} {e^{m_{B^{(-)}}^2/M^2} \over m_{B^{(+)}}+m_{B^{(-)}}} 
\int ds e^{-s/M^2} \left[ m_{B^{(+)}} \mbox{\rm Im} \Pi_1^M(s) - \mbox{\rm Im}
\Pi_2^M(s) \right]~,\nnb \\
\label{nolabel09}
m_{B^{(-)}}^2 \es {\int_{m_Q^2}^{s_0} s ds e^{-s/M^2} \left[  m_{B^{(+)}} \mbox{\rm Im}
\Pi_1^M(s) - \mbox{\rm Im} \Pi_2^M(s) \right] \over
\int_{m_Q^2}^{s_0} ds e^{-s/M^2} \left[  m_{B^{(+)}} \mbox{\rm Im}     
\Pi_1^M(s) - \mbox{\rm Im} \Pi_2^M(s) \right]}~ \nnb.
\eea
Here $\Pi_1^M$ and $\Pi_2^M$ are the invariant functions corresponding to the
structures $\not\!\!p$ and $I$, respectively. The expressions of $\Pi_1^M$ and
$\Pi_2^M$ for the $\Sigma_Q^0$, $\Xi_Q^\prime$ and $\Xi_Q$ baryons are presented
in Appendix C.

\section{Numerical analysis}

This section is devoted to the numerical analysis of the sum rules obtained
in the previous section for the magnetic moments of the diagonal and nondiagonal
transitions of the $J^P={1\over 2}^-$ heavy baryons.
The main input parameters of the light cone QCD sum rules are the photon
distribution amplitudes (DAs). The photon DAs are obtained in \cite{Rsbs27},
and for completeness we present their expressions in Appendix A. Sum rules
for the magnetic moment, together with the photon DAs, also contain the
following input parameters: quark condensate $\qq$, $m_0^2$ that appears in
determination of the vacuum expectation value of the dimension--5 operator
$\la \bar{q} G q \ra = m_0^2 \qq$, magnetic
susceptibility $\chi$ of quarks, etc. In the present analysis we use $\left[ \uu =
\dd \right]_{\mu=1~GeV} = -(0.243)^3~GeV^3$ \cite{Rsbs28}, 
$\sp \ve_{\mu=1~GeV} = 0.8 \uu\ve_{\mu=1~GeV}$, $m_0^2=(0.8\pm 0.2)~GeV^2$
\cite{Rsbs29}. The values of the magnetic susceptibilities can be found in
numerous works (see for example \cite{Rsbs30,Rsbs31,Rsbs32}). In our
numerical analysis we use $\chi(1~GeV)=-2.85~GeV^2$ obtained in
\cite{Rsbs32}, and $\Lambda=(0.5 \pm 0.1)~ GeV$ \cite{Rsbs33}.

Having determined input parameters, in this section we shall proceed with
the analysis of the sum rules for the diagonal and transition
magnetic moments of the $J^P={1\over 2}^-$ heavy baryons.
The sum rules contain three auxiliary parameters:
continuum threshold $s_0$, Borel mass square $M^2$, and $t$ appearing in the
expression of the interpolating current. According to the QCD sum rules
philosophy, any measurable quantity must be independent of these parameters.
For this reason we try to find such regions of these parameters where
magnetic moments are insensitive to their variations. This issue can be
accomplished by the following three-step procedure. First, at fixed values
of $s_0$ and $t$, we try to find region of $M^2$ where magnetic moment is
independent of its variation. The upper bound of $M^2$ is determined by
requiring that the contributions of higher states and continuum constitute,
say, less than 40\% of the contribution coming from the perturbative part.
The lower bound of $M^2$ is obtained by demanding that higher twist
contributions are less than the leading twist contributions. Analysis of our
sum rules leads to the following regions of $M^2$ where magnetic moments are
independent on its variation.
\bea
\label{nolabel06}
&& 2.5~GeV^2 \le M^2 \le 4.0~GeV^2,~\mbox{\rm
for}~\Sigma_c,~\Xi_c^\prime,~\Lambda_c, \Xi_c~, \nnb \\
&& 4.5~GeV^2 \le M^2 \le 7.0~GeV^2,~\mbox{\rm
for}~\Sigma_b,~\Xi_b^\prime,~\Lambda_b, \Xi_b~. \nnb
\eea
Next, we try to the find the domain of variation of the continuum threshold
$s_0$, which is the energy square where the continuum starts. The difference
$\sqrt{s_0}-m$, $m$ being the ground state mass, is the energy needed for the
excitation of the particle to its first excited state, and usually this
difference is varies in the range between $0.3~GeV$ and $0.8~GeV$. In our
analysis we we use the average value $\sqrt{s_0}-m=0.5~GeV$.

As an example, in Figs. (1) and (2)  we present the dependence of the
magnetic moments of $\Sigma_c^0$ baryon, at
$s_0=12~GeV^2$, and magnetic moment of the $\Sigma_b \to
\Lambda_b$ transition at $s_0=40~GeV^2$, on $M^2$, respectively.
It follows from these figures 
that, indeed, we have very good stability of the magnetic moment $\mu$ as
$M^2$ varies in its above-mentioned working region. We also analyze these
dependencies at $s_0=11~GeV^2$ and $s_0=42.5~GeV^2$; and find out that the discrepancy in the
values of the magnetic moment is about 10\%. In other words, the magnetic
moments of  $\Sigma_c^0$ baryon and $\Sigma_b \to \Lambda_b$ transition exhibit the expected
insensitivity to the variations in $s_0$ and $M^2$.

Having decided the working regions of $M^2$ and $s_0$, the third and last
step is to find the working region of the parameter $t$ in which
the predictions for the values magnetic moments of heavy baryons show 
good stability. For this aim, we study the dependence of the
$\Sigma_c^0$ baryon and $\Sigma_b \to \Lambda_b$ transition magnetic moments
on $\cos\theta$,
where $t=\tan\theta$ which are presented in Figs. (3) and (4).
We observe from Figs. (3) and (4)
that, the magnetic moments of diagonal transitions are independent of the
variation in $\cos\theta$ when it varies in the region $-0.7 \le \cos\theta
\le -0.4$; and in the domain $-1.0 \le \cos\theta \le -0.7$ for the
transition magnetic moments, respectively. Our numerical analysis predicts
that the magnetic moment of the
$\Sigma_c^0$ baryon is $\mu = (-2.0\pm 0.1) \mu_N$, and for the $\Sigma_b \to
\Lambda_b$ transition $\mu=(-0.3 \pm 0.05)\mu_N$, where $\mu_N$ is the
nuclear magneton. Performing similar analysis we have calculated the
diagonal and transition magnetic moments of the other $J^P={1\over 2}^-$ heavy baryons whose 
results are presented in Table 2. We note here that, in many
cases, the naive expectation that the relation between the negative and
positive parity baryons, i.e.,
\bea
\label{nolabel07}
\ve \mu_B^- \ve = {m_B^+ \over m_B^-} \ve \mu_B^+ \ve~, \nnb
\eea
is violated considerably. This violation can be attributed to the fact that
in our analysis we take into account contributions coming from
positive-to-positive and nondiagonal transitions.



\begin{table}[!htbp]

\renewcommand{\arraystretch}{1.3}
\addtolength{\arraycolsep}{-0.5pt}
\small
$$
\begin{array}{|c|c||c|c||c|c|}
\hline \hline 
                 & \mu            &                  & \mu           &
                 & \mu                              \\ \hline
\Sigma_b^+       &  1.3  \pm 0.3  & \Sigma_c^{++}    &  2.2 \pm 0.2  &
\Sigma_c^+ \to \Lambda_c^+        & 0.25 \pm 0.05    \\
\Sigma_b^0       &  0.5  \pm 0.05 & \Sigma_c^+       &  0.15\pm 0.02 &
\Xi_c^{\prime 0} \to \Xi_c^0      & 0.08 \pm 0.01    \\
\Sigma_b^-       & -0.3  \pm 0.1  & \Sigma_c^0       & -2.0 \pm 0.1  &
\Xi_c^{\prime +} \to \Xi_c^+      & 0.20 \pm 0.05    \\
\Xi_b^{\prime -} & -0.4  \pm 0.1  & \Xi_c^{\prime 0} & -2.0 \pm 0.2  &
\Xi_b^{\prime 0} \to \Xi_b^0      & -0.008 \pm 0.001 \\
\Xi_b^{\prime 0} &  0.4  \pm 0.1  & \Xi_c^{\prime +} &  0.15\pm 0.02 &
\Xi_b^{\prime -} \to \Xi_b^-      & 0.10 \pm 0.01    \\
\Omega_b^-       & -0.3  \pm 0.1  & \Omega_c^0       & -2.0 \pm 0.2  &
                 &                                   \\
\Lambda_b^0      & -0.11 \pm 0.02 & \Lambda_c^+      &  1.3 \pm 0.2  &
                 &                                   \\
\Xi_b^-          & -0.7  \pm 0.1  & \Xi_c^0          &  1.6 \pm 0.2  &
                 &                                   \\
\Xi_b^0          & -0.12 \pm 0.02 & \Xi_c^+          &  1.2 \pm 0.2  &
                 &                                   \\
\hline \hline 
\end{array}
$$
\caption{Diagonal and transition magnetic moments of the negative
parity, spin-1/2 baryons
containing single heavy quark belonging to the
sextet and antitriplet representations, in units of nuclear magneton
$\mu_N$}
\renewcommand{\arraystretch}{1}
\addtolength{\arraycolsep}{-1.0pt}
\end{table}       

As can easily be seen from Table 2, the transition magnetic moments between
the neutral $\Xi_Q^\prime$ and $\Xi_Q$ baryons are very close to zero. The
magnetic moments of $\Lambda_b^0$, $\Xi_b^0$, $\Sigma_b^+$ and $\Xi_c^+$ are
also small enough to be measured. However the diagonal and transition
magnetic moments of the remaining baryons are quite large, and can be
measured in future experiments.  

In conclusion, the diagonal and nondiagonal transition magnetic moments 
of the negative parity spin-1/2 heavy baryons are calculated in framework of
the LCSR. The contributions coming from the positive to positive, as well as
positive to negative parity transitions are eliminated by constructing
various sum rules It is obtained that the magnetic moments of $\Lambda_b^0$,
$\Xi_b^0$, $\Sigma_c^+$ and $\Xi_c^\prime$ baryons, as well as the
transition magnetic moments between the neutral $\Xi_Q^\prime$ and $\Xi_Q$
baryons are small, while all other magnetic moments are quite large and can
be measured in future.
\newpage
A comparison our results
on the magnetic moments of the negative parity baryons with the predictions
of other approaches, such as quark model, bag model, chiral perturbation
theory, lattice QCD, etc., would be interesting.


\newpage


\section*{Appendix A: Photon distribution amplitudes}
\setcounter{equation}{0}
\setcounter{section}{0}


In this Appendix we present the definitions of the matrix elements of the
form $\la \gamma (q) \ve \bar{q} \Gamma q \ve 0\ra$ in terms of the photon
DAs, and the explicit expressions of the photon DAs entering into the
matrix elements \cite{Rsbs27}.

\bea
\label{esbs14}
&&\langle \gamma(q) \vert  \bar q(x) \sigma_{\mu \nu} q(0) \vert  0
\rangle  = -i e_q \bar q q (\varepsilon_\mu q_\nu - \varepsilon_\nu
q_\mu) \int_0^1 du e^{i \bar u qx} \left(\chi \varphi_\gamma(u) +
\frac{x^2}{16} \mathbb{A}  (u) \right) \nnb \\ &&
-\frac{i}{2(qx)}  e_q \qq \left[x_\nu \left(\varepsilon_\mu - q_\mu
\frac{\varepsilon x}{qx}\right) - x_\mu \left(\varepsilon_\nu -
q_\nu \frac{\varepsilon x}{q x}\right) \right] \int_0^1 du e^{i \bar
u q x} h_\gamma(u)
\nnb \\
&&\langle \gamma(q) \vert  \bar q(x) \gamma_\mu q(0) \vert 0 \rangle
= e_q f_{3 \gamma} \left(\varepsilon_\mu - q_\mu \frac{\varepsilon
x}{q x} \right) \int_0^1 du e^{i \bar u q x} \psi^v(u)
\nnb \\
&&\langle \gamma(q) \vert \bar q(x) \gamma_\mu \gamma_5 q(0) \vert 0
\rangle  = - \frac{1}{4} e_q f_{3 \gamma} \epsilon_{\mu \nu \alpha
\beta } \varepsilon^\nu q^\alpha x^\beta \int_0^1 du e^{i \bar u q
x} \psi^a(u)
\nnb \\
&&\langle \gamma(q) | \bar q(x) g_s G_{\mu \nu} (v x) q(0) \vert 0
\rangle = -i e_q \qq \left(\varepsilon_\mu q_\nu - \varepsilon_\nu
q_\mu \right) \int {\cal D}\alpha_i e^{i (\alpha_{\bar q} + v
\alpha_g) q x} {\cal S}(\alpha_i)
\nnb \\
&&\langle \gamma(q) | \bar q(x) g_s \tilde G_{\mu \nu} i \gamma_5 (v
x) q(0) \vert 0 \rangle = -i e_q \qq \left(\varepsilon_\mu q_\nu -
\varepsilon_\nu q_\mu \right) \int {\cal D}\alpha_i e^{i
(\alpha_{\bar q} + v \alpha_g) q x} \tilde {\cal S}(\alpha_i)
\nnb \\
&&\langle \gamma(q) \vert \bar q(x) g_s \tilde G_{\mu \nu}(v x)
\gamma_\alpha \gamma_5 q(0) \vert 0 \rangle = e_q f_{3 \gamma}
q_\alpha (\varepsilon_\mu q_\nu - \varepsilon_\nu q_\mu) \int {\cal
D}\alpha_i e^{i (\alpha_{\bar q} + v \alpha_g) q x} {\cal
A}(\alpha_i)
\nnb \\
&&\langle \gamma(q) \vert \bar q(x) g_s G_{\mu \nu}(v x) i
\gamma_\alpha q(0) \vert 0 \rangle = e_q f_{3 \gamma} q_\alpha
(\varepsilon_\mu q_\nu - \varepsilon_\nu q_\mu) \int {\cal
D}\alpha_i e^{i (\alpha_{\bar q} + v \alpha_g) q x} {\cal
V}(\alpha_i) \nnb \\ && \langle \gamma(q) \vert \bar q(x)
\sigma_{\alpha \beta} g_s G_{\mu \nu}(v x) q(0) \vert 0 \rangle  =
e_q \qq \left\{
        \left[\left(\varepsilon_\mu - q_\mu \frac{\varepsilon x}{q x}\right)\left(g_{\alpha \nu} -
        \frac{1}{qx} (q_\alpha x_\nu + q_\nu x_\alpha)\right) \right. \right. q_\beta
\nnb \\ && -
         \left(\varepsilon_\mu - q_\mu \frac{\varepsilon x}{q x}\right)\left(g_{\beta \nu} -
        \frac{1}{qx} (q_\beta x_\nu + q_\nu x_\beta)\right) q_\alpha
\nnb \\ && -
         \left(\varepsilon_\nu - q_\nu \frac{\varepsilon x}{q x}\right)\left(g_{\alpha \mu} -
        \frac{1}{qx} (q_\alpha x_\mu + q_\mu x_\alpha)\right) q_\beta
\nnb \\ &&+
         \left. \left(\varepsilon_\nu - q_\nu \frac{\varepsilon x}{q.x}\right)\left( g_{\beta \mu} -
        \frac{1}{qx} (q_\beta x_\mu + q_\mu x_\beta)\right) q_\alpha \right]
   \int {\cal D}\alpha_i e^{i (\alpha_{\bar q} + v \alpha_g) qx} {\cal T}_1(\alpha_i)
\nnb \\ &&+
        \left[\left(\varepsilon_\alpha - q_\alpha \frac{\varepsilon x}{qx}\right)
        \left(g_{\mu \beta} - \frac{1}{qx}(q_\mu x_\beta + q_\beta x_\mu)\right) \right. q_\nu
\nnb \\ &&-
         \left(\varepsilon_\alpha - q_\alpha \frac{\varepsilon x}{qx}\right)
        \left(g_{\nu \beta} - \frac{1}{qx}(q_\nu x_\beta + q_\beta x_\nu)\right)  q_\mu
\nnb \\ && -
         \left(\varepsilon_\beta - q_\beta \frac{\varepsilon x}{qx}\right)
        \left(g_{\mu \alpha} - \frac{1}{qx}(q_\mu x_\alpha + q_\alpha x_\mu)\right) q_\nu
\nnb \\ &&+
         \left. \left(\varepsilon_\beta - q_\beta \frac{\varepsilon x}{qx}\right)
        \left(g_{\nu \alpha} - \frac{1}{qx}(q_\nu x_\alpha + q_\alpha x_\nu) \right) q_\mu
        \right]
    \int {\cal D} \alpha_i e^{i (\alpha_{\bar q} + v \alpha_g) qx} {\cal T}_2(\alpha_i)
\nnb \\ &&+
        \frac{1}{qx} (q_\mu x_\nu - q_\nu x_\mu)
        (\varepsilon_\alpha q_\beta - \varepsilon_\beta q_\alpha)
    \int {\cal D} \alpha_i e^{i (\alpha_{\bar q} + v \alpha_g) qx} {\cal T}_3(\alpha_i)
\nnb \\ &&+
        \left. \frac{1}{qx} (q_\alpha x_\beta - q_\beta x_\alpha)
        (\varepsilon_\mu q_\nu - \varepsilon_\nu q_\mu)
    \int {\cal D} \alpha_i e^{i (\alpha_{\bar q} + v \alpha_g) qx} {\cal T}_4(\alpha_i)
                        \right\}~,
\eea
where $\chi$ is the magnetic susceptibility,
$\varphi_\gamma(u)$ is the leading twist-2, $\psi^v(u)$,
$\psi^a(u)$, ${\cal A}$ and ${\cal V}$ are the twist-3, and
$h_\gamma(u)$, $\mathbb{A}$, and ${\cal T}_i$ ($i=1,~2,~3,~4$) are the
twist-4 photon DAs.
The measure ${\cal D} \alpha_i$ is defined as
\bea
\label{nolabel05}
\int {\cal D} \alpha_i = \int_0^1 d \alpha_{\bar q} \int_0^1 d
\alpha_q \int_0^1 d \alpha_g \delta(1-\alpha_{\bar
q}-\alpha_q-\alpha_g)~.\nnb
\eea

The expressions of the DAs entering into the above matrix elements are
defined as:

\bea
\label{nolabel27}
\varphi_\gamma(u) \es 6 u \bar u \Big[ 1 + \varphi_2(\mu)
C_2^{\frac{3}{2}}(u - \bar u) \Big]~,
\nnb \\
\psi^v(u) \es 3 [3 (2 u - 1)^2 -1 ]+\frac{3}{64} (15
w^V_\gamma - 5 w^A_\gamma)
                        [3 - 30 (2 u - 1)^2 + 35 (2 u -1)^4]~,
\nnb \\
\psi^a(u) \es [1- (2 u -1)^2] [ 5 (2 u -1)^2 -1 ]
\frac{5}{2}
    \Bigg(1 + \frac{9}{16} w^V_\gamma - \frac{3}{16} w^A_\gamma
    \Bigg)~,
\nnb \\
{\cal A}(\alpha_i) \es 360 \alpha_q \alpha_{\bar q} \alpha_g^2
        \Bigg[ 1 + w^A_\gamma \frac{1}{2} (7 \alpha_g - 3)\Bigg]~,
\nnb \\
{\cal V}(\alpha_i) \es 540 w^V_\gamma (\alpha_q - \alpha_{\bar q})
\alpha_q \alpha_{\bar q}
                \alpha_g^2~,
\nnb \\
h_\gamma(u) \es - 10 (1 + 2 \kappa^+ ) C_2^{\frac{1}{2}}(u
- \bar u)~,
\nnb \\
\mathbb{A}(u) \es 40 u^2 \bar u^2 (3 \kappa - \kappa^+ +1 ) +
        8 (\zeta_2^+ - 3 \zeta_2) [u \bar u (2 + 13 u \bar u) + 
                2 u^3 (10 -15 u + 6 u^2) \ln(u) \nnb \\ 
\ar 2 \bar u^3 (10 - 15 \bar u + 6 \bar u^2)
        \ln(\bar u) ]~,
\nnb \\
{\cal T}_1(\alpha_i) \es -120 (3 \zeta_2 + \zeta_2^+)(\alpha_{\bar
q} - \alpha_q)
        \alpha_{\bar q} \alpha_q \alpha_g~,
\nnb \\
{\cal T}_2(\alpha_i) \es 30 \alpha_g^2 (\alpha_{\bar q} - \alpha_q)
    [(\kappa - \kappa^+) + (\zeta_1 - \zeta_1^+)(1 - 2\alpha_g) +
    \zeta_2 (3 - 4 \alpha_g)]~,
\nnb \\
{\cal T}_3(\alpha_i) \es - 120 (3 \zeta_2 - \zeta_2^+)(\alpha_{\bar
q} -\alpha_q)
        \alpha_{\bar q} \alpha_q \alpha_g~,
\nnb \\
{\cal T}_4(\alpha_i) \es 30 \alpha_g^2 (\alpha_{\bar q} - \alpha_q)
    [(\kappa + \kappa^+) + (\zeta_1 + \zeta_1^+)(1 - 2\alpha_g) +
    \zeta_2 (3 - 4 \alpha_g)]~,\nnb \\
{\cal S}(\alpha_i) \es 30\alpha_g^2\{(\kappa +
\kappa^+)(1-\alpha_g)+(\zeta_1 + \zeta_1^+)(1 - \alpha_g)(1 -
2\alpha_g)\nnb \\ 
\ar\zeta_2
[3 (\alpha_{\bar q} - \alpha_q)^2-\alpha_g(1 - \alpha_g)]\}~,\nnb \\
\widetilde {\cal S}(\alpha_i) \es-30\alpha_g^2\{(\kappa -
\kappa^+)(1-\alpha_g)+(\zeta_1 - \zeta_1^+)(1 - \alpha_g)(1 -
2\alpha_g)\nnb \\ 
\ar\zeta_2 [3 (\alpha_{\bar q} -
\alpha_q)^2-\alpha_g(1 - \alpha_g)]\}. \nnb
\eea
The parameters entering  the above DA's are borrowed from
\cite{Rsbs27} whose values are $\varphi_2(1~GeV) = 0$, 
$w^V_\gamma = 3.8 \pm 1.8$, $w^A_\gamma = -2.1 \pm 1.0$, 
$\kappa = 0.2$, $\kappa^+ = 0$, $\zeta_1 = 0.4$, $\zeta_2 = 0.3$, 
$\zeta_1^+ = 0$, and $\zeta_2^+ = 0$.


\newpage


\section*{Appendix B}
In this Appendix we present the expressions of the invariant functions
$\Pi_i$ appearing in
the sum rules for the magnetic moments of $\Sigma_b^0$ baryon, and
$\Xi_b^{\prime 0} \to \Xi_b^0$ transition .\\\\

\setcounter{equation}{0}
\setcounter{section}{0}


{\bf FOR THE MAGNETIC MOMENT OF THE $\Sigma_b^0$ BARYON}\\\\  

{\bf 1) Coefficient of the $(\varepsilon\!\cdot\!p) I$ structure}
%
%
\bea
\Pi_1^{B} \es
-{3\over 128 \pi^4} (-1 + t)^2 (e_b + e_d + e_u) m_b^3 M^3 
  \left( {\cal I}_2 - 2 m_b^2 {\cal I}_3 + m_b^4 {\cal I}_4 \right) \nnb \\
\ar {1\over 1536 \pi^4}
(-1 + t) m_b M^4 \Big\{3 \left[(e_d + e_u) (1 - t) \GG +
48 (1 + t) e_b m_b \pi^2 (\dd + \uu)\right] {\cal I}_2 \nnb \\
\ar 4 m_b^2 \left[ \left(e_d + e_u\right) (-1 + t) \GG + 
      72 m_b (1 + t) \pi^2 \left(e_u \dd +e_d \uu\right) \right]
{\cal I}_3 \Big\} \nnb \\
\ek {3\over 16 \pi^2} (-1 + t^2) m_b^4 M^4 \left (e_d \dd + e_u \uu\right)
\widetilde{j}(h_\gamma) {\cal I}_3 \nnb \\
\ar {1\over 16 \pi^2} (-1 + t)^2 (e_d + e_u) f_{3\gamma} m_b^3
   M^2 \left({\cal I}_2 - m_b^2 {\cal I}_3\right) \psi^v(u_0)\nnb \\
\ar {3\over 32 \pi^2} (-1 + t^2) e_b m_b^2 M^2 (\dd + \uu) {\cal I}_1 \nnb \\
\ar {1\over 768 \pi^2} m_b M^2 {\cal I}_2 \Big\{-3 (-1 + t^2) m_b \left[\dd (7 e_b - 2 e_u) m_0^2 + 
      24 \dd e_b m_b^2 \right. \nnb \\
\ar \left. (7 e_b - 2 e_d) m_0^2 \uu + 24 e_b m_b^2
\uu \right] + 
    2 (-1 + t)^2 (e_d + e_u) f_{3\gamma} \GG \psi^v(u_0)\Big\} \nnb \\
\ek {e^{-m_b^2/M^2}\over 2304 m_b \pi^2}
(-1 + t) M^2 \Big\{9 (1 + t) m_0^2 m_b \left(7 \dd e_b + 12 e_u \dd + 7 e_b \uu + 
      12 e_d \uu\right) \nnb \\
\ar 2 f_{3\gamma} \left[(e_d + e_u) (-1 + t) \GG 
+ 96 (1 + t) m_b \pi^2 \left(e_d \uu + e_u \dd\right)\right] 
     \psi^v(u_0)\Big\} \nnb \\
\ar {e^{-m_b^2/M^2}\over 48 M^2}
(-1 + t^2) f_{3\gamma} m_0^2 m_b^2 \left(e_u \dd + e_d \uu\right)
\psi^v(u_0) \nnb\\
\ar {e^{-m_b^2/M^2}\over 13824 M^4 \pi^2}
(-1 + t^2) \GG m_b^2 \left(e_u \dd + e_d \uu\right) 
  \left[9 m_0^2 + 16 f_{3\gamma} \pi^2 \psi^v(u_0)\right] \nnb \\
\ar {e^{-m_b^2/M^2}\over 1728 M^6}
(-1 + t^2) f_{3\gamma} \GG m_0^2 m_b^2 \left(e_u \dd + e_d \uu\right) \psi^v(u_0) \nnb \\
\ek {e^{-m_b^2/M^2}\over 3456 M^8}  
(-1 + t^2) f_{3\gamma} \GG m_0^2 m_b^4 \left(e_u \dd + e_d \uu\right) \psi^v(u_0) \nnb \\
\ar {1\over 768 \pi^2}
(-1 + t^2) \left[-2 \GG \left(e_u \dd + e_d \uu\right) - 21 e_b m_0^2 m_b^2
\left(\dd + \uu\right) 
     {\cal I}_1\right] \nnb \\
\ar {e^{-m_b^2/M^2}\over 384 \pi^2}
(-1 + t^2) \GG (e_d \dd + e_u \uu) 
   \widetilde{j}(h_\gamma) \nnb \\
\ar {e^{-m_b^2/M^2}\over 96} 
(-1 + t^2) f_{3\gamma} m_0^2 (e_u \dd + e_d \uu) \psi^v(u_0)~. \nnb
\eea
\\\\

{\bf 2) Coefficient of the $(\varepsilon \! \cdot\! p)\!\not\!p$ structure}

\bea
\Pi_2^{B} \es
- {1\over 128 \pi^4} m_b^2 M^6 \Big\{3 (5 + 2 t + 5 t^2) (e_d + e_u) m_b^2 
     \left({\cal I}_3 - 2 m_b^2 {\cal I}_4 + m_b^4 {\cal I}_5\right) \nnb \\
\ar e_b \left[(3 + 2 t + 3 t^2) {\cal I}_2 - 3 (1 + t)^2 m_b^2 {\cal I}_3 - 
      3 (-1 + t)^2 m_b^4 {\cal I}_4 + (3 - 2 t +
3 t^2) m_b^6 {\cal I}_5 \right]\Big\} \nnb \\
\ar {1\over 128 \pi^4}
(3 + 2 t + 3 t^2) e_b m_b^2 M^4 \left(-{\cal I}_1 + 3 m_b^2 {\cal I}_2 - 
   3 m_b^4 {\cal I}_3 + m_b^6 {\cal I}_4\right) \nnb \\
\ar {1\over 1536 \pi^4}
m_b^2 M^2 \Big\{(5 + 2 t + 5 t^2) (e_d + e_u) \GG \nnb \\
\ar 288 (-1 + t^2) m_b \pi^2 \left[ e_u \dd + e_d \uu
+ e_b \left(\dd + \uu\right)\right]\Big\}
\left({\cal I}_2 - m_b^2 {\cal I}_3\right) \nnb \\
\ek {e^{-m_b^2/M^2}\over 768 m_b M^2 \pi^2}
(-1 + t^2) \GG m_0^2 \left(e_u \dd + e_d \uu\right) \nnb \\
\ar {e^{-m_b^2/M^2}\over 1536 M^4 \pi^2}
(-1 + t^2) \GG m_0^2 m_b \left(e_u \dd + e_d \uu\right) \nnb \\
\ek {e^{-m_b^2/M^2}\over 384 m_b \pi^2}
(-1 + t^2) \GG \left(e_u \dd + e_d\uu\right) \nnb \\
\ar {1\over 64 \pi^2}
3 (-1 + t^2) m_0^2 m_b \left (e_u \dd + e_d \uu\right) {\cal I}_1 \nnb \\
\ar {1\over 128 \pi^2} (-1 + t^2) m_b \Big\{\left(e_u \dd + e_d \uu\right) \GG \nnb \\
\ek m_0^2 m_b^2 \left[7 e_b \left(\dd+\uu\right) + 12 \left(e_u \dd + 
 e_d \uu \right) \right] \Big\} {\cal I}_2~.\nnb
\eea
\\\\

{\bf 3) Coefficient of the $\not\!p\!\!\not\!\varepsilon$ structure}

\bea
\Pi_3^{B} \es
{1\over 128 \pi^2}
m_b^2 M^4 \Big\{(-1 + t^2) \left(e_d \dd + e_u \uu\right) 
    \Big[\Big({\cal I}_2 - 2 m_b^2 {\cal I}_3\Big) \Big(i_1({\cal S})
+ i_1(\widetilde{\cal S},1) \nnb \\
\ar i_1({\cal T}_2,1) - 
       i_1({\cal T}_4,1) - 2 i_1({\cal T}_2,v) + 2 i_1({\cal T}_4,v)\Big) + 
     12 m_b^2 {\cal I}_3 \widetilde{j}(h_\gamma)\Big] \nnb \\
\ar (-1 + t)^2 (e_d + e_u) f_{3\gamma} m_b 
    \left(-{\cal I}_2 + m_b^2 {\cal I}_3\right) \psi^{a\prime}(u_0) \Big\} \nnb \\
\ar {e^{-m_b^2/M^2}\over 9216 m_b \pi^2}
(-1 + t)^2 (e_d + e_u) f_{3\gamma} \GG M^2 \psi^{a\prime}(u_0) \nnb \\
\ar {e^{-m_b^2/M^2}\over 96}
(-1 + t^2) f_{3\gamma} M^2 \left(e_u \dd +
e_d \uu\right) \psi^{a\prime}(u_0) \nnb \\
\ek {1\over 3072 \pi^2}
(-1 + t)^2 (e_d + e_u) f_{3\gamma} \GG m_b M^2 {\cal I}_2 
\psi^{a\prime}(u_0) \nnb \\
\ek {e^{-m_b^2/M^2}\over 384 M^2}
(-1 + t^2) f_{3\gamma} m_0^2 m_b^2 \left(e_u \dd + e_d \uu\right) 
   \psi^{a\prime}(u_0) \nnb \\
\ek {e^{-m_b^2/M^2}\over 6912 M^4}
(-1 + t^2) f_{3\gamma} \GG m_b^2 \left(e_u \dd + e_d \uu\right) 
   \psi^{a\prime}(u_0) \nnb \\
\ek {e^{-m_b^2/M^2}\over 13824 M^6}
(-1 + t^2) f_{3\gamma} \GG m_0^2 m_b^2 \left(e_u \dd + e_d \uu\right) 
   \psi^{a\prime}(u_0) \nnb \\
\ar {e^{-m_b^2/M^2}\over 27648 M^8} 
(-1 + t^2) f_{3\gamma} \GG m_0^2 m_b^4 \left(e_u \dd + e_d \uu\right) 
  \psi^{a\prime}(u_0) \nnb \\
\ar {e^{-m_b^2/M^2}\over 9216 \pi^2} 
(-1 + t^2) \Big\{\GG \left(e_d \dd + e_u \uu\right) \Big[i_1({\cal S}) +
i_1(\widetilde{\cal S},1) + i_1({\cal T}_2,1) - 
     i_1({\cal T}_4,1) \nnb \\
\ek 2 i_1({\cal T}_2,v) + 2 i_1({\cal T}_4,v) -
12 \widetilde{j}(h_\gamma)\Big] - 
4 f_{3\gamma} m_0^2 \pi^2 \left(e_u \dd + e_d \uu\right)
\psi^{a\prime}(u_0)\Big\}~.
\eea
\\\\

{\bf 4) Coefficient of the $\not\!\varepsilon$ structure}

\bea
\Pi_4^{B} \es
{1\over 256 \pi^4} 
m_b^2 M^8 \Big\{3 (e_d + e_u) m_b^2 \Big[(1 + t)^2 {\cal I}_3 + 
     4 (1 + t^2) m_b^2 {\cal I}_4 - (5 + 2 t + 5 t^2) m_b^4 {\cal I}_5\Big] \nnb \\ 
\ek e_b \Big[(3 + 2 t + 3 t^2) {\cal I}_2 + 3 (1 + t)^2 m_b^2 {\cal I}_3 - 
     3 (3 + 2 t + 3 t^2) m_b^4 {\cal I}_4 +
(3 - 2 t + 3 t^2) m_b^6 {\cal I}_5\Big]\Big\} \nnb \\
\ar {1\over 256 \pi^4}
e_b m_b^2 M^6 \Big[4 t {\cal I}_1 + 3 m_b^2 {\cal I}_2 - 6 t m_b^2 {\cal I}_2 + 
    3 t^2 m_b^2 {\cal I}_2 - 6 m_b^4 {\cal I}_3 - 6 t^2 m_b^4 {\cal I}_3 + 
    (3 + 2 t + 3 t^2) m_b^6 {\cal I}_4 \Big]\nnb \\
\ar {1\over 128 \pi^2}
(e_d + e_u) f_{3\gamma} m_b^2 M^6 \Big[(1 + t)^2 {\cal I}_2 - 
    2 (1 + 4 t + t^2) m_b^2 {\cal I}_3\Big] i_2({\cal A},v) \nnb \\
\ar {1\over 128 \pi^2}
(e_d + e_u) f_{3\gamma} m_b^2 M^6 \Big[(1 + t)^2 {\cal I}_2 - 
    4 (1 + t + t^2) m_b^2 {\cal I}_3\Big] i_2({\cal V},v) \nnb \\
\ar {1\over 32 \pi^2}
(3 + 2 t + 3 t^2) (e_d + e_u) f_{3\gamma} m_b^4 M^6 {\cal I}_3 \psi^v(u_0)) \nnb \\
\ar {1\over 64 \pi^2}
(-1 + t^2) m_b^3 M^6 \left(e_d \dd + e_u \uu \right) \chi 
   (-{\cal I}_2 + m_b^2 {\cal I}_3) \varphi_\gamma^\prime (u_0) \nnb \\
\ek {1\over 128 \pi^2}  
 (3 + 2 t + 3 t^2) (e_d + e_u) f_{3\gamma} m_b^4 M^6 {\cal I}_3] 
   \psi^{a\prime}(u_0) \nnb \\
\ar {e^{-m_b^2/M^2}\over 4608 m_b \pi^2} 
(-1 + t^2) \GG M^4 \left(e_d \dd + e_u \uu \right) \chi \varphi_\gamma^\prime(u_0) \nnb \\
\ar {1 \over 3072 \pi^4}
m_b^4 M^4 \Big\{-(5 + 2 t + 5 t^2) (e_d + e_u) \GG \nnb \\
\ek 288 (-1 + t^2) m_b \pi^2 \left[\dd (e_b + e_u) + (e_b + e_d) \uu \right] \Big\} {\cal I}_3 \nnb \\
\ar {1 \over 128 \pi^2}
(-1 + t^2) m_b M^4 \left(e_d \dd + e_u \uu\right) 
   \Big[i_1({\cal S},1) - i_1(\widetilde{\cal S},1) + i_1({\cal T}_1,1) \nnb \\
\ek i_1({\cal T}_2,1) + i_1({\cal T}_3,1) - i_1({\cal T}_4,1)\Big]  {\cal I}_1 \nnb \\
\ar {1 \over 1536 \pi^4}
 m_b M^4 \Big\{
   m_b \Big[(1 + t^2) (e_d + e_u) \GG + 
      144 (-1 + t^2) e_b m_b \pi^2 \left(\dd + \uu\right)\Big] \nnb \\
\ek (-1 + t^2) \pi^2 
     \left(e_d \dd + e_u \uu\right) \Big[6 m_b^2 \Big(2 i_1({\cal S},1) - 2 (i_1(\widetilde{\cal S},1) - 2 i_1({\cal T}_1,1) + 
          i_1({\cal T}_2,1) + i_1({\cal T}_4,1) \nnb \\
\ek 6 i_1({\cal S},v) - 2 i_1({\cal T}_3,v) + 
          2 i_1({\cal T}_4,v)) - \mathbb{A}^\prime (u_0)\Big) + 
      \GG \chi \varphi_\gamma^\prime(u_0)\Big]\Big\} {\cal I}_2 \nnb \\
\ek {e^{-m_b^2/M^2}\over 4608 m_b \pi^2}
(-1 + t^2) \GG M^2 \left(e_d \dd + e_u \uu\right) \Big[i_1({\cal T}_1,1) - i_1({\cal T}_3,1) \nnb \\ 
\ar 6 i_1({\cal S},v) + 2 i_1({\cal T}_3,v) - 2 i_1({\cal T}_4,v)\Big] \nnb \\
\ar {e^{-m_b^2/M^2}\over 9216 \pi^2}
(e_d + e_u) f_{3\gamma} \GG M^2 \Big[(1 + 6 t + t^2) i_2({\cal A},v) + 
   (3 + 2 t + 3 t^2) i_2({\cal V},v)\Big] \nnb \\
\ar {e^{-m_b^2/M^2}\over 2304 \pi^2}
f_{3\gamma} M^2 \Big[-(3 + 2 t + 3 t^2) (e_d + e_u) \GG - 
    288 (-1 + t^2)  m_b \pi^2 \left(e_u \dd + e_d \uu\right) \Big] 
   \psi^v(u_0) \nnb \\
\ar {1\over 256 \pi^2}
(-1 + t^2) m_b M^2 \Big\{e_u \dd \GG - \dd (7 e_b + 12 e_u) m_0^2 m_b^2 \nnb \\
\ar \Big[e_d \GG - (7 e_b + 12 e_d) m_0^2 m_b^2\Big] \uu \Big] {\cal I}_2 \nnb \\
\ar {e^{-m_b^2/M^2}\over 9216 m_b \pi^2}
 (-1 + t^2) M^2 \Big[-36 \left(\GG - 6 m_0^2 m_b^2\right) \left( e_u \dd + e_d \uu\right) \nnb \\ 
\ar \GG \left(e_d \dd + e_u \uu\right) \mathbb{A}^\prime (u_0)\Big] \nnb \\
\ar {e^{-m_b^2/M^2}\over 9216 \pi^2} 
 f_{3\gamma} M^2 \Big[(3 + 2 t + 3 t^2) (e_d + e_u) \GG +
     288 (-1 + t^2) m_b \pi^2 \left(e_u \dd + e_d \uu\right)\Big] \psi^{a\prime}(u_0) \nnb \\
\ar {e^{-m_b^2/M^2}\over 768 M^2 \pi^2}
(-1 + t^2) m_b \left(e_u \dd + e_d \uu\right) 
  \Big[\GG m_0^2 + \pi^2 f_{3\gamma} \left(\GG - 6 m_0^2 m_b^2\right)\nnb \\
\cp \left(-4 \psi^v(u_0) + \psi^{a\prime}(u_0)\right)\Big]] \nnb \\
\ar {e^{-m_b^2/M^2}\over 9216 M^4 \pi^2}
(-1 + t^2) \GG m_b \left(e_u \dd + e_d \uu\right) \Big[-3 m_0^2 m_b^2 \nnb \\
\ek 2 \pi^2 f_{3\gamma} (3 m_0^2 - 2 m_b^2) \left(4 \psi^v(u_0) - 
     \psi^{a\prime}(u_0)\right)\Big] \nnb \\
\ar {e^{-m_b^2/M^2}\over 1536 M^6}
(-1 + t^2) f_{3\gamma} \GG m_0^2 m_b^3 \left(e_u \dd + e_d \uu\right) 
  \left(4 \psi^v(u_0) - \psi^{a\prime}(u_0)\right) \nnb \\
\ek {e^{-m_b^2/M^2}\over 9216 M^8}
(-1 + t^2) f_{3\gamma} \GG m_0^2 m_b^5 \left(e_u \dd + e_d \uu\right)
   \left(4 \psi^v(u_0) - \psi^{a\prime}(u_0)\right) \nnb \\
\ar {e^{-m_b^2/M^2}\over 18432 \pi^2}
(-1 + t^2) \GG m_b \left(\dd e_d + e_u \uu\right) \Big[2 i_1({\cal T}_1,1) - 2 i_1({\cal T}_3,1) + 
    12 i_1({\cal S},v) \nnb \\
\ar 4 i_1({\cal T}_3,v) - 4 i_1({\cal T}_4,v) - 
    \mathbb{A}^\prime(u_0)\Big] \nnb \\
\ek {e^{-m_b^2/M^2}\over 1536 m_b \pi^2} 
 (-1 + t^2) \left(e_u \dd + e_d \uu\right) \Big[\GG (m_0^2 - 2 m_b^2) + 
    12 f_{3\gamma} m_0^2 m_b^2 \pi^2 \left(4 \psi^v(u_0) -
\psi^{a\prime}(u_0)\right)\Big]~. \nnb
\eea

\newpage

{\bf FOR THE $\Xi_b^{\prime 0}\to \Xi_b^0$ TRANSITION MAGNETIC MOMENT}\\\\

{\bf 1) Coefficient of the $(\varepsilon\!\cdot\!p) I$ structure}

%
%
\bea
\Pi_1^B \es
- {1 \over 32 \sqrt{3} \pi^2}
(-1 + t) m_b^2 M^4 \Big\{4 (2 + t) m_b^2 \left(e_u \sp - e_s \uu\right) {\cal I}_3 \nnb \\
\ar e_b \left(\sp - \uu\right) \Big[(7 + 3 t) {\cal I}_2 - 2 (3 + t) 
m_b^2 {\cal I}_3\Big]\Big\} \nnb \\
\ar {\sqrt{3} \over 8 \pi^2}
(-1 + t) m_b^4 M^4 \left(e_s \sp - e_u \uu\right) 
   {\cal I}_3 \widetilde{j}(h_\gamma) \nnb \\
\ar {1 \over 16 \sqrt{3} \pi^2} 
 (-1 + t) (3 + t) (e_s - e_u) f_{3\gamma} m_b^3 M^4 
   \left({\cal I}_2 - m_b^2 {\cal I}_3\right) \psi^v(u_0) \nnb \\
\ar {e^{-m_b^2/M^2}\over 768 \sqrt{3} \pi^2}
(-1 + t) M^2 \Big\{12 m_0^2 \left(e_u \sp - e_s \uu\right) 
     \Big[4 + t \left(2 + m_b^2 e^{m_b^2/M^2} {\cal I}_2\right)\Big] \nnb \\
\ar e_b \left(\sp - \uu\right) 
     \Big[ 24 (7 + 3 t) m_b^2 e^{m_b^2/M^2} \left(-{\cal I}_1 + m_b^2 {\cal I}_2\right) + 
      m_0^2 \Big(7 (1 + t) + (29 + 17 t) m_b^2 e^{m_b^2/M^2} {\cal I}_2\Big)\Big]\Big\} \nnb \\
\ar {e^{-m_b^2/M^2}\over 1152 \sqrt{3} m_b \pi^2}
(-1 + t) f_{3\gamma} M^2 
\Big\{e_u \Big[-96 (1 + t) m_b \pi^2 \sp - (3 + t) \GG 
       \left(-1 + 3 m_b^2 e^{m_b^2/M^2} {\cal I}_2\right)\Big] \nnb \\
\ar e_s \Big[96 (1 + t) m_b \pi^2 \uu + 
      (3 + t) \GG \left(-1 + 3 m_b^2 e^{m_b^2/M^2} {\cal I}_2\right)\Big]\Big\} \psi^v(u_0) \nnb \\
\ar {e^{-m_b^2/M^2}\over 48 \sqrt{3} M^2}
(-1 + t^2) f_{3\gamma} m_0^2 m_b^2 \left(e_u \sp - e_s \uu\right) \psi^v(u_0) \nnb \\
\ar {e^{-m_b^2/M^2}\over 6912 \sqrt{3} M^4 \pi^2}
(-1 + t) \GG m_b^2 \left(e_u \sp - e_s \uu\right) \Big[-3 (2 + t) m_0^2 + 
   8 (1 + t) f_{3\gamma} \pi^2 \psi^v(u_0)\Big] \nnb \\
\ar {e^{-m_b^2/M^2}\over 1728 \sqrt{3} M^6}
(-1 + t^2) f_{3\gamma} \GG m_0^2 m_b^2 \left(e_u \sp - e_s \uu\right) \psi^v(u_0) \nnb \\
\ek {e^{-m_b^2/M^2}\over 3456 \sqrt{3} M^8}
(-1 + t^2) f_{3\gamma} \GG m_0^2 m_b^4 \left(e_u \sp - e_s \uu\right) \psi^v(u_0) \nnb \\
\ar {e^{-m_b^2/M^2}\over 2304 \sqrt{3} \pi^2} 
(-1 + t) \Big[4 (2 + t) \GG \left(e_u \sp - e_s \uu\right) + 
    3 (29 + 17 t) e_b m_0^2 m_b^2 e^{m_b^2/M^2} \left(\sp - \uu\right) {\cal I}_1\Big] \nnb \\
\ek {e^{-m_b^2/M^2}\over 192 \sqrt{3} \pi^2}
(-1 + t) \GG \left(e_s \sp - e_u \uu\right) 
   \widetilde{j}(h_\gamma) \nnb \\
\ek {e^{-m_b^2/M^2}\over 288 \sqrt{3}}
(-1 + t^2) f_{3\gamma} m_0^2 \left(e_u \sp - e_s \uu\right) \psi^v(u_0)~. \nnb
\eea
\\\\\\

{\bf 2) Coefficient of the $(\varepsilon \! \cdot\! p)\!\not\!p$ structure}

\bea
\Pi_2^B \es
{1\over 8 \sqrt{3} \pi^2}
(-2 + t + t^2) m_b^3 M^2 \Big[(e_b + e_u) \sp - (e_b + e_s) \uu\Big] 
  \left({\cal I}_2 - m_b^2 {\cal I}_3\right) \nnb \\
\ek {e^{-m_b^2/M^2}\over 1152 \sqrt{3} m_b M^2 \pi^2}
(-2 + t + t^2) \GG m_0^2 \left(e_u \sp - e_s \uu\right) \nnb \\
\ar {e^{-m_b^2/M^2}\over 2304 \sqrt{3} M^4 \pi^2}
(-2 + t + t^2) \GG m_0^2 m_b \left(e_u \sp - e_s \uu\right) \nnb \\
\ek {e^{-m_b^2/M^2}\over 576 \sqrt{3} m_b \pi^2}
(-1 + t) (2 + t) \GG \left(e_u \sp - e_s \uu\right) \nnb \\
\ar {\sqrt{3}\over 64 \pi^2}
(-1 + t^2) m_0^2 m_b \left(e_u \sp - e_s \uu\right) {\cal I}_1 \nnb \\
\ar {1\over 192 \sqrt{3} \pi^2}
 (-1 + t) m_b \Big\{\Big[(2 + t) e_u \GG - 
      3 (3 + 2 t) e_b m_0^2 m_b^2 -3 (7 + 5 t) e_u m_0^2 m_b^2 \Big] \sp \nnb \\
\ar \Big[-(2 + t) e_s \GG + 3 (3 + 2 t) e_b m_0^2 m_b^2 + 3 (7 + 5 t) e_s m_0^2 
       m_b^2\Big] \uu\Big\} {\cal I}_2~. \nnb
\eea
\\\\


{\bf 3) Coefficient of the $\not\!p\!\!\not\!\varepsilon$ structure}

\bea
\Pi_3^B \es
{1 \over 256 \sqrt{3} \pi^4}
(-1 + t) m_b^3 M^6 \Big[ -3 (3 + t) (e_s - e_u) 
    \left({\cal I}_2 - 2 m_b^2 {\cal I}_3 + m_b^4 {\cal I}_4\right) \nnb \\
\ar 8 (-1 + t) m_b \pi^2 
    \left(e_s \sp - e_u \uu\right) \chi \left({\cal I}_3 - m_b^2 {\cal
I}_4\right) \varphi_\gamma^\prime (u_0)\Big] \nnb \\
\ek {1 \over 768 \sqrt{3} \pi^4}
(-1 + t) (3 + t) m_b^3 M^4 \Big[ \GG (e_u - e_s) + 
    24 (e_b - e_u) m_b \pi^2 \sp \nnb \\
\ar  24 (-e_b + e_s) m_b \pi^2 \uu \Big]
   {\cal I}_3 \nnb \\
\ar {1 \over 1024 \sqrt{3} \pi^4}
(e_s - e_u) m_b M^4 \Big\{-(-1 + t) (3 + t) \GG {\cal I}_2 + 
   16 f_{3\gamma} m_b^2 \pi^2 \left(-{\cal I}_2
+ m_b^2 {\cal I}_3\right) \nnb \\ 
\cp    \Big[2 (-1 + t) (3 + t) \psi^v(u_0) - (-1 + t^2) 
      \psi^{a\prime}(u_0)\Big] \Big\} \nnb \\
\ar {1 \over 128 \sqrt{3} \pi^2}
(-1 + t) m_b^2 M^4 \left(e_s \sp - e_u \uu\right) {\cal I}_2 
   \Big\{(5 + t) i_1({\cal S},1) + (1 + 5 t) i_1(\widetilde{\cal S},1) \nnb \\
\ar 2 i_1({\cal T}_1,1) + i_1({\cal T}_2,1) + 
    2 i_1({\cal T}_3,1) - 5 i_1({\cal T}_4,1) - 6 i_1({\cal S},v)
 - 2 i_1(\widetilde{\cal S},v) \nnb \\
\ek t \Big[2 i_1({\cal T}_1,1) - 5 i_1({\cal T}_2,1) + 2 i_1({\cal T}_3,1) + i_1({\cal T}_4,1) + 2 i_1({\cal S},v) + 
      6 i_1(\widetilde{\cal S},v) \nnb \\
\ar 4 i_1({\cal T}_2,v)- 4 i_1({\cal T}_3,v)\Big] - 
    4 i_1({\cal T}_3,v) + 4 i_1({\cal T}_4,v)\Big\} \nnb \\
\ek {1 \over 128 \sqrt{3} \pi^2} 
(-1 + t) m_b^4 M^4 \left(e_s \sp - e_u \uu\right) {\cal I}_3 
   \Big\{4 (2 + t) i_1({\cal S},1) + (4 + 8 t) i_1(\widetilde{\cal S},1) \nnb \\
\ek 4 \Big[(-1 + t) i_1({\cal T}_1,1) - i_1({\cal T}_2,1) + 2 i_1({\cal T}_4,1) + 3 i_1({\cal S},v) + 
      i_1(\widetilde{\cal S},v) + i_1({\cal T}_2,v) \nnb \\
\ar t \Big(-2 i_1({\cal T}_2,1) + i_1({\cal T}_4,1) + 
        i_1({\cal S},v) + 3 i_1(\widetilde{\cal S},v)
+ i_1({\cal T}_2,v)\Big)\Big] \nnb \\
\ar 4 (1 + t) i_1({\cal T}_4,v) + 8 (2 + t) \widetilde{j}(h_\gamma) + 
    (-1 + t) \mathbb{A}^\prime (u_0)\Big\} \nnb \\
\ar {e^{-m_b^2/M^2}\over 768 \sqrt{3} \pi^2}
(-1 + t) M^2 \Big\{m_0^2 \left(e_u \sp - e_s \uu\right) \Big[-6 (3 + t) + 
      (7 + t) m_b^2 e^{m_b^2/M^2} {\cal I}_2\Big] \nnb \\
\ar e_b \left(\sp - \uu\right)
\Big[(11 + 5 t) m_0^2 - 24 (3 + t) m_b^2 e^{m_b^2/M^2} \left(-{\cal I}_1 +
m_b^2 {\cal I}_2\right)\Big]\Big\} \nnb \\
\ek {e^{-m_b^2/M^2}\over 2304 \sqrt{3} m_b \pi^2}
(-1 + t) (3 + t) f_{3\gamma} M^2 
   \Big[-(e_s - e_u) \GG - 96 m_b \pi^2 \left(e_u \sp 
- e_s \uu \right) \nnb \\
\ar 3 (e_s - e_u) \GG m_b^2 e^{m_b^2/M^2} {\cal I}_2\Big] \psi^v(u_0) \nnb \\
\ek {1 \over 2304 \sqrt{3} \pi^2}
(-1 + t)^2 \GG m_b^2 M^2 \left(e_s \sp - e_u \uu\right) \chi {\cal I}_2 
   \varphi_\gamma^\prime (u_0) \nnb \\
\ar {e^{-m_b^2/M^2}\over 4608 \sqrt{3} m_b \pi^2}
f_{3\gamma} M^2 \Big[96 t (-1 + t) m_b \pi^2 \left(e_u \sp -
e_s \uu\right) \nnb \\
\ar (-1 + t^2) (e_s - e_u) \GG \left(-1 + 3 m_b^2 e^{m_b^2/M^2} {\cal I}_2\right)\Big] 
  \psi^{a\prime}(u_0) \nnb \\
\ek {e^{-m_b^2/M^2}\over 192 \sqrt{3} M^2}
(-1 + t) f_{3\gamma} m_0^2 m_b^2 \left(e_u \sp - e_s \uu\right) 
   \Big[2 (3 + t) \psi^v(u_0) + t \psi^{a\prime}(u_0)\Big] \nnb \\
\ar {e^{-m_b^2/M^2}\over 27648 \sqrt{3} M^4 \pi^2}
(-1 + t) \GG m_b^2 \left(e_u \sp - e_s \uu\right) \Big\{3 (3 + t) m_0^2 \nnb \\
\ek 8 f_{3\gamma} \pi^2 \Big[2 (3 + t) \psi^v(u_0) + 
     t \psi^{a\prime}(u_0)\Big]\Big\} \nnb \\
\ek {e^{-m_b^2/M^2}\over 6912 \sqrt{3} M^6}
(-1 + t) f_{3\gamma} \GG m_0^2 m_b^2 \left(e_u \sp - e_s \uu\right) 
   \Big[2 (3 + t) \psi^v(u_0) + t \psi^{a\prime}(u_0)\Big] \nnb \\
\ar {e^{-m_b^2/M^2}\over 13824 \sqrt{3} M^8}
(-1 + t) f_{3\gamma} \GG m_0^2 m_b^4 \left(e_u \sp - e_s \uu\right) 
  \Big[2 (3 + t) \psi^v(u_0) + t \psi^{a\prime}(u_0)\Big] \nnb \\
\ar {e^{-m_b^2/M^2}\over 9216 \sqrt{3} \pi^2}
(-1 + t) \GG \left(e_s \sp - e_u \uu\right) \Big\{3 (1 + t) i_1({\cal S},1) + 
   3 (1 + t) i_1(\widetilde{\cal S},1) \nnb \\
\ar 2 i_1({\cal T}_1,1) + 3 i_1({\cal T}_2,1) - 2 i_1({\cal T}_3,1) - 
   3 i_1({\cal T}_4,1) - 6 i_1({\cal S},v) - 2 i_1(\widetilde{\cal S},v) - 4 i_1({\cal T}_2,v) \nnb \\
\ar 4 i_1({\cal T}_3,v) + 16 \widetilde{j}(h_\gamma) - \mathbb{A}^\prime (u_0) + 
   t \Big[-2 i_1({\cal T}_1,1) + 3 i_1({\cal T}_2,1) + 2 i_1({\cal T}_3,1) - 3 i_1({\cal T}_4,1) \nnb \\
\ek 2 i_1({\cal S},v) - 6 i_1(\widetilde{\cal S},v) - 4 i_1({\cal T}_3,v) + 4 i_1({\cal T}_4,v) + 
     8 \widetilde{j}(h_\gamma) + \mathbb{A}^\prime (u_0)\Big]\Big\} \nnb \\
\ek {e^{-m_b^2/M^2}\over 2304 \sqrt{3} \pi^2}
(-1 + t) \Big\{(3 + t) \GG \left(e_u \sp - e_s \uu\right) + 
    3 (11 + 5 t) e_b m_0^2 m_b^2 e^{m_b^2/M^2} \left(\sp - \uu\right) {\cal I}_1 \nnb \\
\ar 2 f_{3\gamma} m_0^2 \pi^2 \left(e_u \sp - e_s \uu\right) \Big[2 (11 + 5 t) \psi^v(u_0) + 
      (2 + 5 t) \psi^{a\prime}(u_0)\Big]\Big\}~. \nnb
\eea
\\\\\\


{\bf 4) Coefficient of the $\not\!\varepsilon$ structure}

\bea
\Pi_4^B \es
{\sqrt{3}\over 32 \pi^4}
(1 + t + t^2) (e_s - e_u) m_b^4 M^8 
  \left(-{\cal I}_3 + m_b^2 {\cal I}_4\right) \nnb \\
\ar {1\over 128 \sqrt{3} \pi^2}
(e_s - e_u) f_{3\gamma} m_b^2 M^6 \left[-3 (1 + t)^2 {\cal I}_2 + 
    4 (1 + t + t^2) m_b^2 {\cal I}_3\right] i_2({\cal A},v) \nnb \\
\ar {1\over 128 \sqrt{3} \pi^2}
(e_s - e_u) f_{3\gamma} m_b^2 M^6 \left[-3 (1 + t)^2 {\cal I}_2 + 
    2 (1 + 4 t + t^2) m_b^2 {\cal I}_3\right] i_2({\cal V},v) \nnb \\
\ek {1\over 32 \sqrt{3} \pi^2}
(1 + t + t^2) (e_s - e_u) f_{3\gamma} m_b^4 M^6 {\cal I}_3
\left[4 \psi^v(u_0)-\psi^{a\prime}(u_0)\right] \nnb \\
\ar {\sqrt{3}\over 64 \pi^2}
(-1 + t^2) m_b^3 M^6 \left (e_s \sp - e_u \uu\right) \chi 
   \left({\cal I}_2 - m_b^2 {\cal I}_3\right) \varphi_\gamma^\prime (u_0) \nnb \\
\ar {e^{-m_b^2/M^2}\over 1536 \sqrt{3} m_b \pi^2}
(-1 + t^2) \GG M^4 \left(e_s \sp - e_u \uu\right)
\chi \left(-1 + 3 m_b^2 e^{m_b^2/M^2} {\cal I}_2\right)
\varphi_\gamma^\prime (u_0) \nnb \\
\ek {1 \over 16 \sqrt{3} \pi^2}
(-2 + t + t^2) m_b^5 M^4 \left[(e_b + e_u) \sp - (e_b + e_s)
\uu\right] {\cal I}_3 \nnb \\
\ek {\sqrt{3} \over 128 \pi^2}
(-1 + t^2) m_b M^4 \left(e_s \sp - e_u \uu\right) {\cal I}_1 
    \left[i_1({\cal T}_1,1) + i_1({\cal T}_3,1)\right] \nnb \\
\ek {1 \over 128 \sqrt{3} \pi^2}
(-1 + t) m_b M^4 \left(e_s \sp - e_u \uu\right) {\cal I}_1 
   \Big\{(5 + t) i_1({\cal S},1) \nnb \\
\ek (1 + 5 t) \left[i_1(\widetilde{\cal S},1) + i_1({\cal T}_2,1)\right] - 
    (5 + t) i_1({\cal T}_4,1)\Big\} \nnb \\
\ek {1 \over 768 \sqrt{3} \pi^4}
m_b^2 M^4 {\cal I}_2 \Big\{-(1 + t + t^2) e_s \GG + 
    (1 + t + t^2) e_u \GG \nnb \\
\ek 48 (-2 + t + t^2) e_b m_b \pi^2 
     \left(\sp - \uu\right) + 3 (-1 + t) m_b \pi^2 \left(e_s \sp - e_u \uu\right) \nnb \\
\cp \Big[2 (-1 + 7 t) \left(i_1(\widetilde{\cal S},1) + i_1({\cal T}_2,1)\right) + 
      2 (-7 + t) \left(i_1({\cal S},1) - i_1({\cal T}_4,1)\right) \nnb\\
\ar 8 (2 + t) i_1({\cal S},v) - 
      4 (-1 + t) i_1(\widetilde{\cal S},v) + 8 t i_1({\cal T}_4,v) + 
      4 (3 + t) \left(i_1({\cal T}_2,v) - 2 \widetilde{j}(h_\gamma)\right) \nnb \\
\ar 3 (1 + t) \left(-4 (i_1({\cal T}_1,1) + i_1({\cal T}_3,v)) + \mathbb{A}^\prime
      (u_0)\right)\Big]\Big\} \nnb \\
\ek {e^{-m_b^2/M^2}\over 9216 \sqrt{3} \pi^2}
(-1 + t)^2 (e_s - e_u) f_{3\gamma} \GG M^2 \left[i_2({\cal A},v) -
i_2({\cal V},v)\right] \nnb \\
\ek {e^{-m_b^2/M^2}\over 9216 \sqrt{3} m_b \pi^2}
(-1 + t) \GG M^2 \left(e_s \sp - e_u \uu\right) \Big\{4 (-1 + t) i_1({\cal S},1) \nnb \\
\ar 2 \Big[2 (-1 + t) i_1(\widetilde{\cal S},1) - 3 (1 + t) i_1({\cal T}_1,1) - 2 i_1({\cal T}_2,1) + 
      3 i_1({\cal T}_3,1) \nnb \\
\ar  2 \Big(i_1({\cal T}_4,1) + 4 i_1({\cal S},v) + i_1(\widetilde{\cal S},v) + 
        3 i_1({\cal T}_2,v) - 3 i_1({\cal T}_3,v) - 6 \widetilde{j}(h_\gamma) \Big) \nnb \\
\ar t \Big(2 i_1({\cal T}_2,1) + 3 i_1({\cal T}_3,1) - 2 i_1({\cal T}_4,1)
+ 4 i_1({\cal S},v) - 2 i_1(\widetilde{\cal S},v) + 2 i_1({\cal T}_2,v) \nnb \\
\ek 6 i_1({\cal T}_3,v) + 4  i_1({\cal T}_4,v) - 
          4 \widetilde{j}(h_\gamma)\Big)\Big] + 3 (1 + t) \mathbb{A}^\prime (u_0)\Big\} \nnb \\
\ek {1 \over 128 \sqrt{3} \pi^2}
(-1 + t) (7 + 5 t) m_0^2 m_b^3 M^2 \left(e_u \sp - e_s \uu\right) {\cal I}_2 \nnb \\
\ek {e^{-m_b^2/M^2}\over 384 \sqrt{3} m_b \pi^2}
(-2 + t + t^2) M^2 \left(e_u \sp - e_s \uu\right) \Big[\GG \left(1 -
e^{m_b^2/M^2} m_b^2 {\cal I}_2\right) \nnb \\
\ek 6 m_0^2 m_b^2 + 32 f_{3\gamma} m_b^2 \pi^2 \psi^v(u_0)\Big] \nnb \\
\ar {e^{-m_b^2/M^2}\over 2304 \sqrt{3} \pi^2}
(1 + t + t^2) (e_s - e_u) f_{3\gamma} \GG M^2 
\left[4 \psi^v(u_0) - \psi^{a\prime}(u_0)\right] \nnb \\
\ar {e^{-m_b^2/M^2}\over 48 \sqrt{3}}
(1 + t - 2 t^2) e_s f_{3\gamma} m_b M^2 \uu \psi^{a\prime}(u_0) \nnb \\
\ar {e^{-m_b^2/M^2}\over 384 \sqrt{3} \pi^2} 
(-1 + t) m_b M^2 \Big[- e^{m_b^2/M^2}3 (3 + 2 t) e_b m_0^2 m_b^2 
\left(\sp - \uu\right) {\cal I}_2 \nnb \\
\ar 8 (1 + 2 t) e_u f_{3\gamma} \pi^2 \sp \psi^{a\prime}(u_0)\Big] \nnb \\
\ar {e^{-m_b^2/M^2}\over 1152 \sqrt{3} M^2 \pi^2}
(-1 + t) m_b \left(e_u \sp - e_s \uu\right) \Big\{(2 + t) \GG m_0^2 \nnb \\
\ar f_{3\gamma} \left(\GG - 6 m_0^2 m_b^2\right) \pi^2 \Big[-4 (2 + t) \psi^v(u_0) + 
     (1 + 2 t) \psi^{a\prime}(u_0)\Big]\Big\} \nnb \\
\ek {e^{-m_b^2/M^2}\over 13824 \sqrt{3} M^4 \pi^2}
(-1 + t) \GG m_b \left(e_u \sp - e_s \uu\right) \Big\{3 (2 + t) m_0^2 m_b^2 \nnb \\
\ar    2 f_{3\gamma} (3 m_0^2 - 2 m_b^2) \pi^2 \Big[4 (2 + t) \psi^v(u_0) - 
      (1 + 2 t) \psi^{a\prime}(u_0)\Big]\Big\} \nnb \\
\ar {e^{-m_b^2/M^2}\over 2304 \sqrt{3} M^6}
(-1 + t) f_{3\gamma} \GG m_0^2 m_b^3 \left(e_u \sp - e_s \uu\right) 
  \Big[4 (2 + t) \psi^v(u_0) - (1 + 2 t) \psi^{a\prime}(u_0)\Big] \nnb \\
\ek {e^{-m_b^2/M^2}\over 13824 \sqrt{3} M^8}
(-1 + t) f_{3\gamma} \GG m_0^2 m_b^5 \left(e_u \sp - e_s \uu\right) 
   \Big[4 (2 + t) \psi^v(u_0) - (1 + 2 t) \psi^{a\prime}(u_0)\Big] \nnb \\
\ar {e^{-m_b^2/M^2}\over 18432 \sqrt{3} \pi^2}
(-1 + t) \GG m_b \left(e_s \sp - e_u \uu\right) \Big\{4 (-1 + t) i_1({\cal S},1) \nnb \\
\ar 2 \Big[2 (-1 + t) i_1(\widetilde{\cal S},1) - 3 (1 + t) i_1({\cal T}_1,1)
- 2 i_1({\cal T}_2,1) + 3 i_1({\cal T}_3,1) + 2 i_1({\cal T}_4,1) +
8 i_1({\cal S},v) \nnb \\
\ar 2 i_1(\widetilde{\cal S},v) + 6 i_1({\cal T}_2,v) - 6 i_1({\cal T}_3,v)
- 12 \widetilde{j}(h_\gamma) + 
t \Big(2 i_1({\cal T}_2,1) + 3 i_1({\cal T}_3,1) - 2 i_1({\cal T}_4,1)\nnb \\
\ar 4 i_1({\cal S},v) - 2 i_1(\widetilde{\cal S},v) + 2 i_1({\cal T}_2,v)
- 6 i_1({\cal T}_3,v) + 4 i_1({\cal T}_4,v) - 4 \widetilde{j}(h_\gamma)\Big)\Big]
+ 3 (1 + t) \mathbb{A}^\prime (u_0)\Big\} \nnb \\
\ar {e^{-m_b^2/M^2}\over 2304 \sqrt{3} m_b \pi^2} 
\left(e_u \sp - e_s \uu\right) \Big\{-(-2 + t + t^2) \GG (m_0^2 - 2 m_b^2) \nnb \\
\ar 18 (-1 + t^2) f_{3\gamma} m_0^2 m_b^2 \pi^2 \Big[-4 \psi^v(u_0) + 
      \psi^{a\prime}(u_0)\Big]\Big\}~. \nnb
\eea


The functions $i_n~(n=1,2)$, and $\widetilde{j}_1(f(u))$
appearing in the invariant functions above are defined as:
\bea
\label{nolabel}
i_1(\phi,f(v)) \es \int {\cal D}\alpha_i \int_0^1 dv 
\phi(\alpha_{\bar{q}},\alpha_q,\alpha_g) f(v) \delta^\prime(k-u_0)~, \nnb \\
i_2(\phi,f(v)) \es \int {\cal D}\alpha_i \int_0^1 dv 
\phi(\alpha_{\bar{q}},\alpha_q,\alpha_g) f(v) \delta^{\prime\prime}(k-u_0)~, \nnb \\
\widetilde{j}(f(u)) \es \int_{u_0}^1 du f(u)~, \nnb \\
{\cal I}_n \es \int_{m_b^2}^{\infty} ds\, {e^{-s/M^2} \over s^n}~,\nnb
\eea
where 
\bea
k = \alpha_q + \alpha_g \bar{v}~,~~~~~u_0={M_1^2 \over M_1^2
+M_2^2}~,~~~~~M^2={M_1^2 M_2^2 \over M_1^2 +M_2^2}~.\nnb
\eea




\newpage


\section*{Appendix C} 

In this appendix we give the 
expressions of the invariant amplitudes $\Pi_1^M$ and
$\Pi_2^M$ entering into the mass sum rule for the $\Sigma_b^0$ and
$\Xi_b^\prime$ or $\Xi_b$ baryons.
Here the masses of the light quarks are neglected. \\\\

{\bf A) FOR THE $\Sigma_b^0$ BARYON}

\setcounter{equation}{0}
\setcounter{section}{0}

%
%
\bea
&&\Pi_1^M =
{3\over 256 \pi^4} \Big\{ - m_b^4  M^6  [5 + t (2 + 5 t)] \left[
m_b^4 {\cal I}_5 -
2 m_b^2 {\cal I}_4 +
{\cal I}_3\right] \Big\} \nnb \\
\ar {1\over 192 \pi^4} m_b^4 M^2 \left[\GG  (1 + t + t^2 ) - 18 m_b \pi^2 (-1+t^2)
\left(\dd+\uu\right)\right] {\cal I}_3 \nnb \\
\ar {1\over 3072 \pi^4} m_b^2 M^2 \left[-\GG  (13 + 10 t + 13 t^2 ) + 288 m_b \pi^2
(-1+t^2) \left(\dd+\uu\right)\right] {\cal I}_2 \nnb \\
\ar {e^{-m_b^2/M^2} \over 73728 m_b M^2 \pi^4} \Big\{ - \GG^2 m_b (1 + t)^2 +
768 m_b m_0^2 \dd \uu \pi^4 (-1 + t)^2 \nnb \\
\ek 56 \GG m_0^2 \pi^2 (-1 + t^2) \left(\dd + \uu\right) \Big\} \nnb \\
\ar {1 \over 768 M^2 \pi^2} \GG m_b (-1 + t^2) (\dd + \uu) {\cal I}_1 \nnb \\
\ar {e^{-m_b^2/M^2} \over 18432 M^4  \pi^2} m_b m_0^2
\left[\GG \left(\uu+\dd\right) (-1+t^2) +
384 m_b \dd \uu \pi^2 (-1+t)^2 \right] \nnb \\
\ar {e^{-m_b^2/M^2} \over 1728 M^6} m_b^2 \GG (-1 + t)^2 \dd \uu \nnb \\
\ar {e^{-m_b^2/M^2} \over 1728 M^8} m_b^2 m_0^2 \GG (-1 + t)^2 \dd \uu \nnb \\
\ek {e^{-m_b^2/M^2} \over 3456 M^{10}} m_b^4 m_0^2 \GG (-1 + t)^2 \dd \uu \nnb \\
\ek {e^{-m_b^2/M^2} \over 768 m_b \pi^2} \left[ \GG \left(\uu+\dd\right)
(-1+t^2) + 32 m_b \dd \uu \pi^2 (-1+t)^2 \right] \nnb \\
\ar {1\over 256 \pi^2} m_b \left(\uu+\dd\right) (-1+t^2) \left[
\left(\GG - 13 m_b^2 m_0^2\right) {\cal I}_2 + 6 m_0^2 {\cal I}_1
\right]~, \nnb \\ \nnb \\ \nnb \\
%
%
&&\Pi_2^M =
-{3\over 256 \pi^4} \Big\{ - m_b^3  M^6 (-1+t)^2 \left[          
m_b^4 {\cal I}_4 -                          
2 m_b^2 {\cal I}_3 +                               
{\cal I}_2\right] \Big\} \nnb \\       
\ar {1\over 3072 \pi^4} m_b M^4 \Big\{4 m_b^2 \left[\GG (-1+t)^2 + 72 m_b
\left(\uu+\dd\right) \pi^2 (-1+t^2) \right] {\cal I}_3
- 3 \GG (-1+t)^2 {\cal I}_2 \Big\} \nnb \\
\ek {7 e^{-m_b^2/M^2} \over 256 \pi^2} m_0^2 M^2
\left(\uu+\dd\right) (-1 + t^2) \nnb \\
\ar {1\over 1024 \pi^4} m_b M^2 \Big\{
m_b \left[ 3 m_b \GG (-1+t)^2 + 4 m_0^2 \left(\uu+\dd\right) \pi^2 (-1+t^2)
\right] {\cal I}_2 -
2 \GG (-1+t)^2 {\cal I}_1 \Big\} \nnb \\
\ek {e^{-m_b^2/M^2} \over 73728 M^2 \pi^4} m_b 
\left[ \GG^2 (-1 + t)^2 +
1536 m_0^2 \dd \uu \pi^4 (3 + 2 t + 3 t^2) \right] \nnb \\
\ar {e^{-m_b^2/M^2} \over 18432 M^4  \pi^2} m_b 
\left[ -11 m_b m_0^2 \GG \left(\uu+\dd\right) (-1+t^2)\right. \nnb \\
\ek \left. 32 \left(\GG - 12 m_0^2 m_b^2 \right) \dd \uu (5 + 2 t + 5 t^2) 
\right] \nnb \\
\ar {e^{-m_b^2/M^2} \over 1728 M^6} m_b \left( m_b^2 - 3 m_0^2 \right)
\GG \dd \uu (5 + 2 t + 5 t^2) \nnb \\
\ar {e^{-m_b^2/M^2} \over 576 M^8} m_b^3 m_0^2 \GG \dd \uu (5 + 2 t + 5 t^2)
\nnb \\
\ek {e^{-m_b^2/M^2} \over 3456 M^{10}} m_b^5 m_0^2 \GG \dd \uu (5 + 2 t + 5
t^2) \nnb \\
\ar {e^{-m_b^2/M^2} \over 36864 m_b \pi^4} \left[ \GG^2 (-1+t)^2
- 1536 m_b^2 \dd \uu \pi^4 (5 + 2 t + 5 t^2) \right. \nnb \\
\ar \left.96 m_b \GG \left(\uu+\dd\right) \pi^2 (-1+t^2) \right]~.
\eea

\vskip 2 cm

{\bf B) FOR THE $\Xi_b^{\prime 0}$ BARYON}

%
%
\bea
&&\Pi_1^M =
{3\over 256 \pi^4} \Big\{ - m_b^4  M^6  [5 + t (2 + 5 t)] \left[
m_b^4 {\cal I}_5 -
2 m_b^2 {\cal I}_4 +
{\cal I}_3\right] \Big\} \nnb \\
\ar {1\over 192 \pi^4} m_b^4 M^2 \left[\GG  (1 + t + t^2 ) - 18 m_b \pi^2 (-1+t^2)
\left(\sp+\uu\right)\right] {\cal I}_3 \nnb \\
\ar {1\over 3072 \pi^4} m_b^2 M^2 \left[-\GG  (13 + 10 t + 13 t^2 ) + 288 m_b \pi^2
(-1+t^2) \left(\sp+\uu\right)\right] {\cal I}_2 \nnb \\
\ar {e^{-m_b^2/M^2} \over 73728 m_b M^2 \pi^4} \Big\{ - \GG^2 m_b (1 + t)^2 +
768 m_b m_0^2 \sp \uu \pi^4 (-1 + t)^2 \nnb \\
\ek 56 \GG m_0^2 \pi^2 (-1 + t^2) \left(\sp + \uu\right) \Big\} \nnb \\
\ar {1 \over 768 M^2 \pi^2} \GG m_b (-1 + t^2) (\sp + \uu) {\cal  E}_1 \nnb \\
\ar {e^{-m_b^2/M^2} \over 18432 M^4  \pi^2} m_b m_0^2
\left[\GG \left(\uu+\sp\right) (-1+t^2) +
384 m_b \sp \uu \pi^2 (-1+t)^2 \right] \nnb \\
\ar {e^{-m_b^2/M^2} \over 1728 M^6} m_b^2 \GG (-1 + t)^2 \sp \uu \nnb \\
\ar {e^{-m_b^2/M^2} \over 1728 M^8} m_b^2 m_0^2 \GG (-1 + t)^2 \sp \uu \nnb \\
\ek {e^{-m_b^2/M^2} \over 3456 M^{10}} m_b^4 m_0^2 \GG (-1 + t)^2 \sp \uu \nnb \\
\ek {e^{-m_b^2/M^2} \over 768 m_b \pi^2} \left[ \GG \left(\uu+\sp\right)
(-1+t^2) + 32 m_b \sp \uu \pi^2 (-1+t)^2 \right] \nnb \\
\ar {1\over 256 \pi^2} m_b \left(\uu+\sp\right) (-1+t^2) \left[
\left(\GG - 13 m_b^2 m_0^2\right) {\cal I}_2 + 6 m_0^2 {\cal I}_1
\right]~,\nnb
\eea
\\\\
%
%
\bea
&&\Pi_2^M =
-{3\over 256 \pi^4} \Big\{ - m_b^3  M^6 (-1+t)^2 \left[          
m_b^4 {\cal I}_4 -                          
2 m_b^2 {\cal I}_3 +                               
{\cal I}_2\right] \Big\} \nnb \\       
\ar {1\over 3072 \pi^4} m_b M^4 \Big\{4 m_b^2 \left[\GG (-1+t)^2 + 72 m_b
\left(\uu+\sp\right) \pi^2 (-1+t^2) \right] {\cal I}_3
- 3 \GG (-1+t)^2 {\cal I}_2 \Big\} \nnb \\
\ek {7 e^{-m_b^2/M^2} \over 256 \pi^2} m_0^2 M^2
\left(\uu+\sp\right) (-1 + t^2) \nnb \\
\ar {1\over 1024 \pi^4} m_b M^2 \Big\{
m_b \left[ 3 m_b \GG (-1+t)^2 + 4 m_0^2 \left(\uu+\sp\right) \pi^2 (-1+t^2)
\right] {\cal I}_2 -
2 \GG (-1+t)^2 {\cal I}_1 \Big\} \nnb \\
\ek {e^{-m_b^2/M^2} \over 73728 M^2 \pi^4} m_b 
\left[ \GG^2 (-1 + t)^2 +
1536 m_0^2 \sp \uu \pi^4 (3 + 2 t + 3 t^2) \right] \nnb \\
\ar {e^{-m_b^2/M^2} \over 18432 M^4  \pi^2} m_b 
\left[ -11 m_b m_0^2 \GG \left(\uu+\sp\right) (-1+t^2)\right. \nnb \\
\ek \left. 32 \left(\GG - 12 m_0^2 m_b^2 \right) \sp \uu (5 + 2 t + 5 t^2) 
\right] \nnb \\
\ar {e^{-m_b^2/M^2} \over 1728 M^6} m_b \left( m_b^2 - 3 m_0^2 \right)
\GG \sp \uu (5 + 2 t + 5 t^2) \nnb \\
\ar {e^{-m_b^2/M^2} \over 576 M^8} m_b^3 m_0^2 \GG \sp \uu (5 + 2 t + 5 t^2)
\nnb \\
\ek {e^{-m_b^2/M^2} \over 3456 M^{10}} m_b^5 m_0^2 \GG \sp \uu (5 + 2 t + 5
t^2) \nnb \\
\ar {e^{-m_b^2/M^2} \over 36864 m_b \pi^4} \left[ \GG^2 (-1+t)^2
- 1536 m_b^2 \sp \uu \pi^4 (5 + 2 t + 5 t^2) \right. \nnb \\
\ar \left.96 m_b \GG \left(\uu+\sp\right) \pi^2 (-1+t^2) \right]~. \nnb
\eea

\vskip 2 cm

{\bf C) FOR THE $\Xi_b^{0}$ BARYON}
%
%
\bea
&&\Pi_1^M =
- {1\over 256 \pi^4}
(-3 m_b^4 M^6 (5 + 2 t + 5 t^2) \left({\cal I}_3 - 2 m_b^2 {\cal I}_4 + m_b^4 {\cal I}_5\right) \nnb \\
\ar {1\over 3072 \pi^4}
m_b^2 M^2 \Big[3 \GG (1 + t)^2 {\cal I}_2 - 16 \GG m_b^2 (1 + t + t^2) {\cal I}_3 \nnb \\
\ek 32 m_b \pi^2 (-1 + t) (1 + 5 t) \left(\sp + \uu\right) 
\left(-{\cal I}_2 + m_b^2 {\cal I}_3\right)\Big] \nnb \\
\ar {e^{-m_b^2/M^2} \over 221184 m_b M^2 \pi^4} 
\GG^2 m_b (13 + 10 t + 13 t^2) + 768 m_0^2 m_b \pi^4 (-1 + t) (25 + 23 t) \sp \uu \nnb \\
\ek 8 \GG \pi^2 (-1 + t) (\sp + \uu) \left[m_0^2 (1 + 5 t) + 
    12 m_b^2 e^{m_b^2/M^2} (5 + t) {\cal I}_1\right] \nnb \\
\ar {e^{-m_b^2/M^2} \over 55296 M^4 \pi^2}
m_0^2 m_b (-1 + t) \left[384 m_b \pi^2 (13 + 11 t) \sp \uu + 
   \GG (31 + 11 t) \left(\sp + \uu\right)\right] \nnb \\
\ar {e^{-m_b^2/M^2} \over 5184 M^6}
\GG m_b^2 (-1 + t) (13 + 11 t) \sp \uu \nnb \\
\ar {e^{-m_b^2/M^2} \over 5184 M^8}
\GG m_0^2 m_b^2 (-1 + t) (13 + 11 t) \sp \uu \nnb \\
\ek {e^{-m_b^2/M^2} \over 10368 M^{10}}
\GG m_0^2 m_b^4 (-1 + t) (13 + 11 t) \sp \uu \nnb \\
\ar {e^{-m_b^2/M^2} \over 6912 m_b \pi^2}
(-1 + t) \Big\{\GG (1 + 5 t) \left(\sp + \uu\right) 
\left(-1 + 3 m_b^2 e^{m_b^2/M^2} {\cal I}_2\right) \nnb \\
\ek 3 m_b \Big[32 \pi^2 (13 + 11 t) \sp \uu + 3 m_0^2 m_b e^{m_b^2/M^2} \left(\sp + \uu\right) 
      [-6 (1 + t) {\cal I}_1 + m_b^2 (7 + 11 t) 
{\cal I}_2]\Big]\Big\}~.\nnb
\eea
\\\\
%
%
\bea
&&\Pi_2^M =
- {1\over 256 \pi^4}
m_b^3 M^6 (-1 + t) (13 + 11 t) \left({\cal I}_2 - 2 m_b^2 {\cal I}_3 +
m_b^4 {\cal I}_4\right) \nnb \\
\ek {1\over 9216 \pi^4}
m_b M^4 (-1 + t) \Big[-96 m_b^3 \pi^2 (1 + 5 t) (\sp + \uu) {\cal I}_3 + 
    \GG (13 + 11 t) \left(3 {\cal I}_2 - 4 m_b^2 {\cal I}_3\right)\Big] \nnb \\
\ek {e^{-m_b^2/M^2} \over 3072 \pi^4}
M^2 (-1 + t) \Big\{4 m_0^2 \pi^2 \left(\sp + \uu\right) \left[1 + 5 t +
m_b^2 e^{m_b^2/M^2} (5 + t) {\cal I}_2\right] \nnb \\
\ar  \GG m_b e^{m_b^2/M^2} \left[-2 (-1 + t) {\cal I}_1 +
3 m_b^2 (3 + 5 t) {\cal I}_2\right]\Big\} \nnb \\
\ar {e^{-m_b^2/M^2} \over 221184 M^2 \pi^4}
m_b (-1 + t) \Big[\GG^2 (11 + 13 t) - 1536 m_0^2 \pi^4 (-1 + t) \sp \uu \Big] \nnb \\
\ar {e^{-m_b^2/M^2} \over 55296 M^4 \pi^2}
m_b \Big\{1152 m_0^2 m_b^2 \pi^2 (5 + 2 t + 5 t^2) \sp \uu \nnb \\
\ek \GG \Big[96 \pi^2 (5 + 2 t + 5 t^2) \sp \uu + m_0^2 m_b (-1 + t) (29 + t) 
      \left(\sp + \uu\right)\Big]\Big\} \nnb \\
\ar {e^{-m_b^2/M^2} \over 1728 M^6}
\GG m_b (-3 m_0^2 + m_b^2) (5 + 2 t + 5 t^2) \sp \uu \nnb \\
\ar {e^{-m_b^2/M^2} \over 576 M^8}
\GG m_0^2 m_b^3 (5 + 2 t + 5 t^2) \sp \uu \nnb \\
\ek {e^{-m_b^2/M^2} \over 3456 M^{10}}
\GG m_0^2 m_b^5 (5 + 2 t + 5 t^2) \sp \uu \nnb \\
\ar {e^{-m_b^2/M^2} \over 110592 m_b \pi^4}
\Big[\GG^2 (11 + 2 t - 13 t^2) - 4608 m_b^2 \pi^4 (5 + 2 t + 5 t^2) \sp \uu \nnb \\
\ek 32 \GG m_b \pi^2 (-7 + t) (-1 + t) \left(\sp + \uu\right)\Big]~.\nnb
\eea
where
\bea
{\cal I}_n = \int_{m_b^2}^{\infty} ds\, {e^{-s/M^2} \over s^n}~.\nnb
\eea

\newpage

\section*{Figure captions}
{\bf Fig. (1)} The dependence of the magnetic moment of the $\Sigma_c^0$ 
baryon on $M^2$, at several fixed values of $t$, and at
$s_0=12.0~GeV^2$, in units of nuclear magneton $\mu_N$.\\\\
{\bf Fig. (2)} The same as Fig. (1), but for the $\Sigma_b\to \Lambda_b$
transition at $s_0=40.0~GeV^2$.\\\\
{\bf Fig. (3)} The dependence of the magnetic moment of the $\Sigma_c^0$
baryon on $\cos\theta$, at several fixed values of $M^2$, and at
$s_0=12.0~GeV^2$, in units of nuclear magneton $\mu_N$.\\\\
{\bf Fig. (4)} The same as Fig. (3), but for the $\Sigma_b\to \Lambda_b$
transition at $s_0=40.0~GeV^2$.
\newpage

\begin{figure}
\vskip 3. cm
    \includegraphics{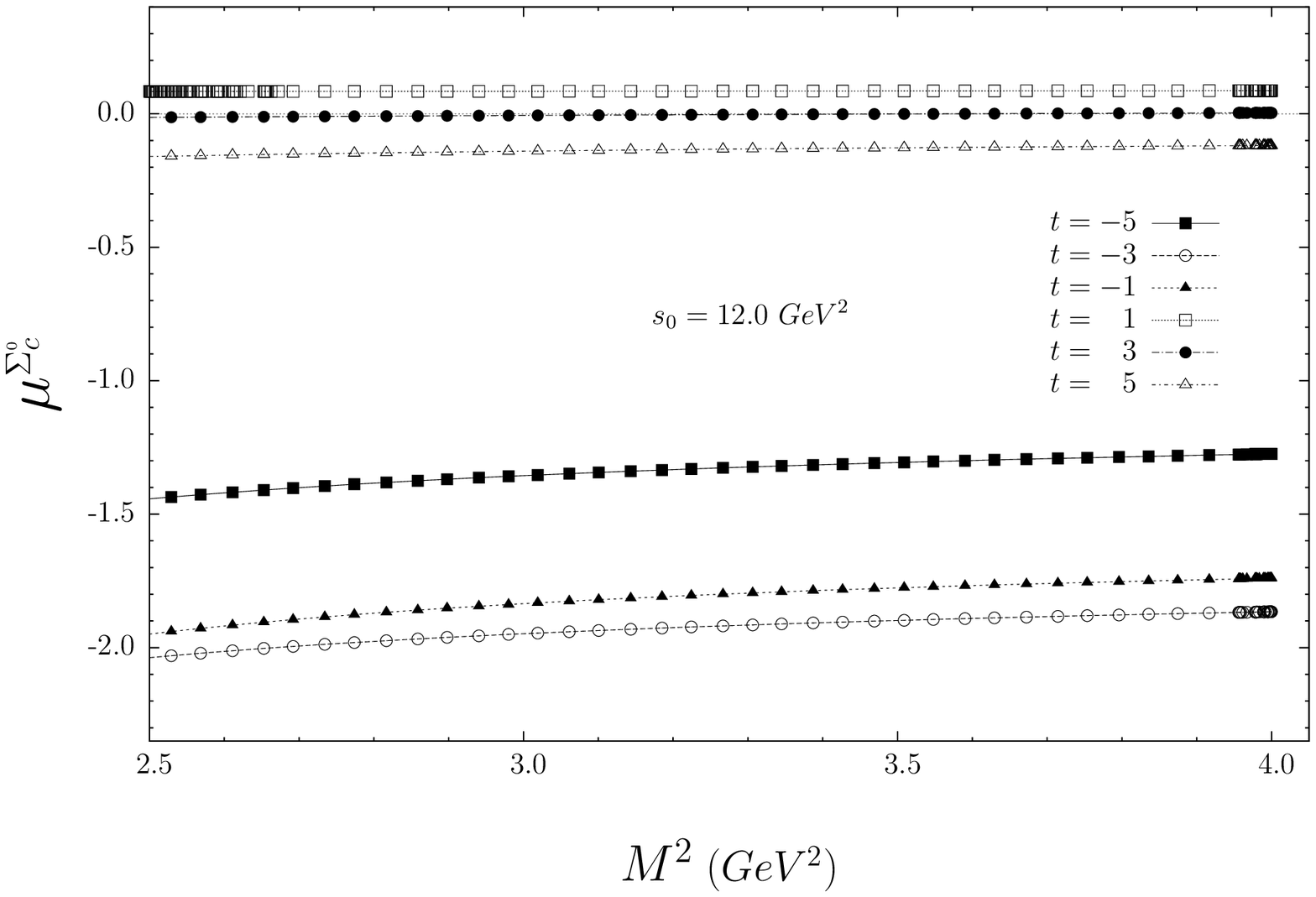}
\vskip 7.0cm
\caption{}
\end{figure}

\begin{figure}
\vskip 3. cm
    \includegraphics{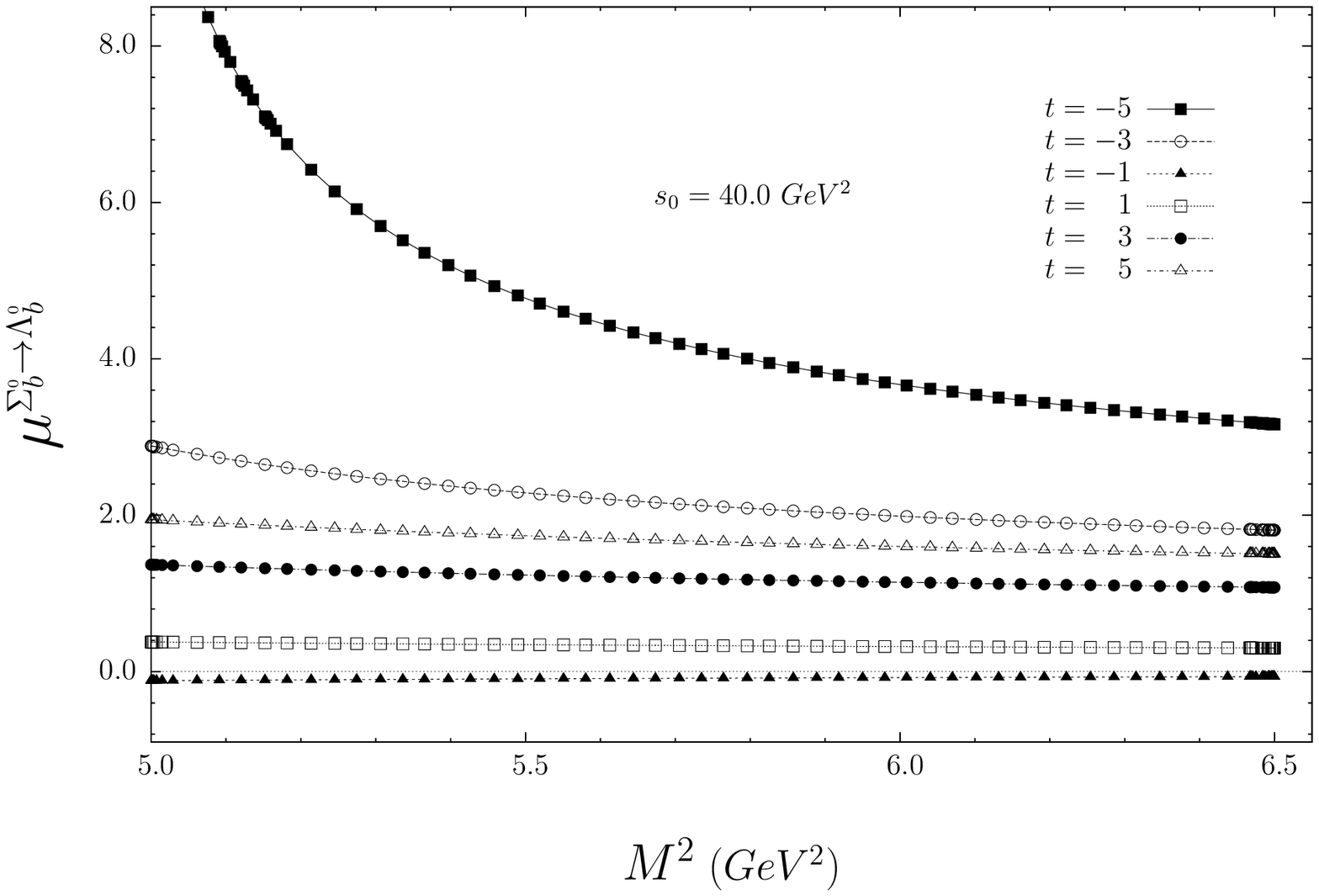}
\vskip 7.0cm
\caption{}
\end{figure}

\begin{figure}
\vskip 3. cm
    \includegraphics{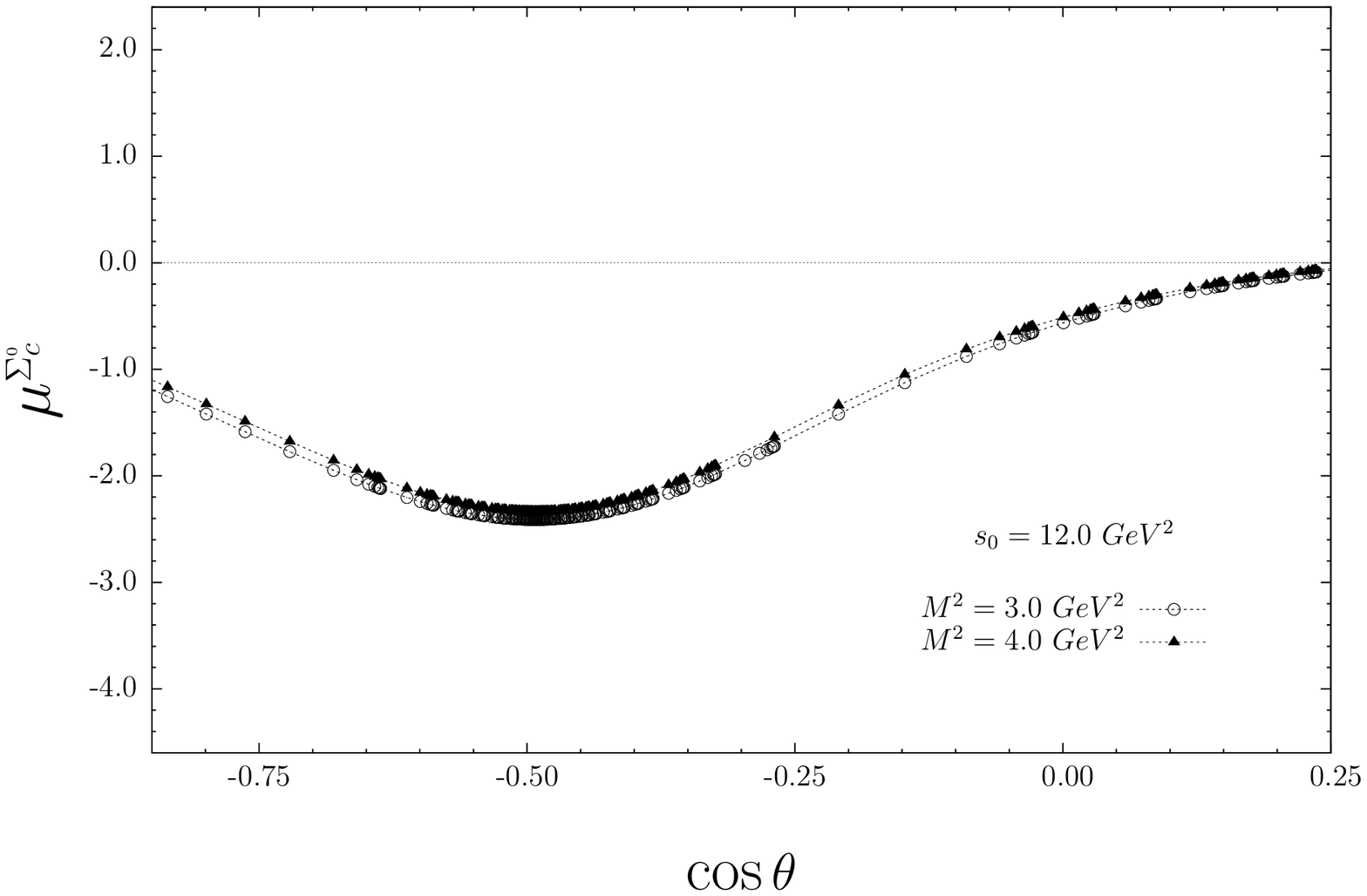}
\vskip 7.0cm
\caption{}
\end{figure}

\begin{figure}
\vskip 3. cm
    \includegraphics{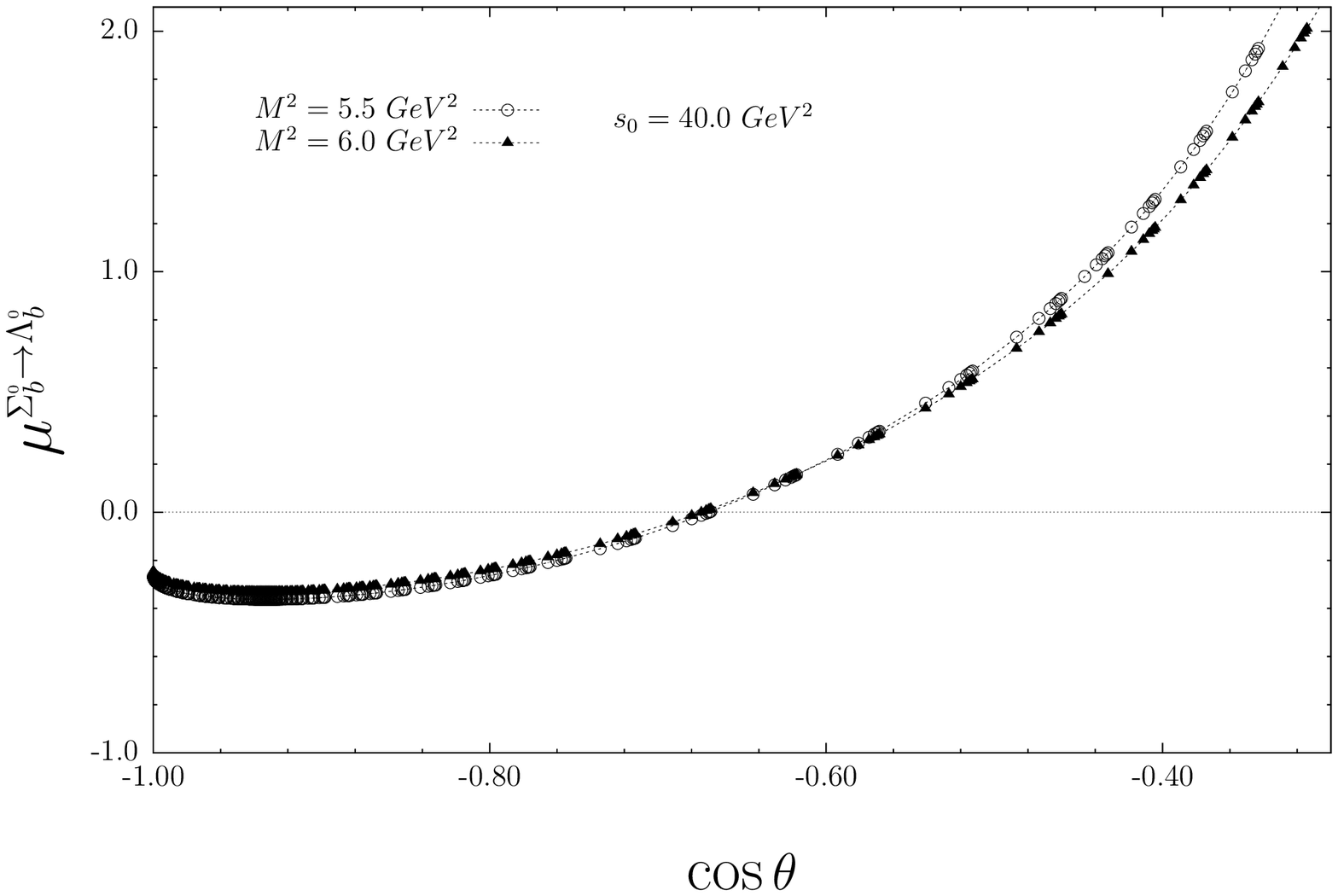}
\vskip 7.0cm
\caption{}
\end{figure}


\newpage

\begin{thebibliography}{99}


\bibitem{Rsbs01} K. Olive {\it et. al}, Particle Data Group,
  Chin. Phys. C {\bf 38}, 090001 (2014).

\bibitem{Rsbs02} R. Muzik {\it et. al}, Belle Collaboration,
  Phys. Rev. Lett. {\bf 94}, 122002 (2005).

\bibitem{Rsbs03} W. Roberts, M. Pervin,
  Int. J. Mod. Phys. A {\bf 23}, 2817 (2008).

\bibitem{Rsbs04} P. G. Ortega, D. R. Entem, and F. Fernandez,
  Phys. Lett. B {\bf 718}, 1381 (2013).
 
\bibitem{Rsbs05} Z. S. Brown, W. Detmold, S. Meinel, and K. Orginos,
 Phys. Rev. D {\bf 79}, 014504 (2009).

\bibitem{Rsbs06} E. Bagan, P. Ball, V. M. Braun, and H. G. Dosch,
  Phys. Lett. B {\bf 278}, 457 (1992);
                 M. Neubert,
  Phys. Rev. D {\bf 45}, 2451 (1992);
                 D, J, Broudhurst, and A. A. Grozin,
  Phys. Lett. B {\bf 274}, 421 (1992);
                 P. Ball, and V. M. Braun,
  Phys. Rev. D {\bf 49}, 2472 (1994);  
                 F. S. Navarra, M. Nielsen,
  Phys. Lett. B {\bf 443}, 285 (1998).

\bibitem{Rsbs07} . E. Klempt, and J. M. Richard
  Rev. Mod. Phys. {\bf 82}, 1095 (2010).

\bibitem{Rsbs08} A. L. Choudry, V. Joshi,
  Phys. Rev. D {\bf 13}, 3115 (1976).

\bibitem{Rsbs09} D. B. Lichtenberg,
  Phys. Rev. D {\bf 15}, 345 (1977).

\bibitem{Rsbs10} B. Julia-Diaz, and D. O. Riska,
  Nucl. Phys. A {\bf 739 }, 69 (2004).

\bibitem{Rsbs11} A. Faessler, T. Gutsche, M. A. Ivanov, J. G. Korner, and V. E. Lyubovitskij,
  Phys. Rev. D {\bf 73}, 094013 (2006).

\bibitem{Rsbs12} C. Albertus, E. Hernandez, J. Nieves, J. M. Verde-Velasco,
  Eur. Phys. J. A {\bf 32}, 183 (2007).

\bibitem{Rsbs13} S. Kumar, R. Dhir, and R. C. Verma,
  J. Phys. G {\bf 31}, 141 (2005).

\bibitem{Rsbs14} R. Dhir, C. S. Kim, R. C. Verma,
  Phys. Rev. D {\bf 88}, 094002 (2013).

\bibitem{Rsbs15} B. Patel, A. K. Rai, P.C. Vinodkumar,
  J. Phys. G {\bf 35}, 065001 (2008).

\bibitem{Rsbs16} N. Sharma, P. K. Chatley, and M. Gupta,
  Phys. Rev. D {\bf 81}, 073001 (2010).

\bibitem{Rsbs17} M. Savage,
  Phys. Lett. B {\bf 326}, 303 (1994).

\bibitem{Rsbs18} A. Bernotas, and V. Simonis,
  arXiv:1209.2900 [hep-ph].

\bibitem{Rsbs19} S. L. Zhu, W. Y. Hwang, and Z. S. Yang,
  Phys. Rev. D {\bf 56}, 7273 (1997).

\bibitem{Rsbs20} T. M. Aliev, A. \"{O}zpineci, M. Savc{\i},
  Phys. Rev. D {\bf 65}, 096004 (2002).

\bibitem{Rsbs21} T. M. Aliev, K. Azizi, M. Savc{\i},
  Phys. Rev. D {\bf 77}, 114006 (2008). 
 
\bibitem{Rsbs22} T. M. Aliev, K. Azizi, A. \"{O}zpineci,
  Nucl. Phys. B {\bf 808}, 137 (2009). 

\bibitem{Rsbs23} V. M. Braun,
  e-Print: hep-ph/9801222.

\bibitem{Rsbs24} E. Bagan, M. Chabab, H. G. Dosch, and S. Narison,
  Phys. Lett. B {\bf 278}, 369 (1992).

\bibitem{Rsbs25} I. I. Balitsky, V. M. Braun,
  Nucl. Phys. B {\bf 311}, 541 (1998).

\bibitem{Rsbs26} V. M. Braun, J. E. Filyanov,
  Z. Phys. C {\bf 48}, 239 (1990).

\bibitem{Rsbs27} P. Ball, V. M. Braun, and N. Kivel,
  Nucl. Phys. B {\bf 649}, 263 (2003).

\bibitem{Rsbs28} B. L. Ioffe,
   Prog. Part. Nucl. Phys. {\bf 56}, 232 (2006).

\bibitem{Rsbs29} V. M. Belyaev, B. L. Ioffe,
  JETP {\bf 56}, 493 (1982).

\bibitem{Rsbs30} V. M. Belyaev, and Y. I. Kogan,
  Yad. Fiz. {\bf 40}, 1035 (1984).

\bibitem{Rsbs31} I. I. Balitsky, A. V. Kolesnichenko, A. V. Yung,
  Yad. Fiz. {\bf 41}, 282 (1985).

\bibitem{Rsbs32} J. Rohrwild,
  J. High Energy Phys. {\bf 09}, 073 (2007).     

\bibitem{Rsbs33} K. G. Chetyrkin, A. Khodjamirian, and A. A. Pivovarov,
  Phys. Lett. B {\bf 661}, 250 (2008).

\end{thebibliography}
\end{document}